\documentclass[12pt]{article}

\usepackage{newtxtext,newtxmath}

\usepackage{graphicx}

\usepackage[letterpaper,margin=1in]{geometry}

\renewenvironment{abstract}
	{\quotation}
	{\endquotation}

\date{}

\makeatletter
\renewcommand{\fnum@figure}{\textbf{Figure \thefigure}}
\renewcommand{\fnum@table}{\textbf{Table \thetable}}
\makeatother

\usepackage{scicite}

\usepackage{url}

\def\scititle{
	The microscopic origin of the Quantum Hall Effect
}
\title{\bfseries \boldmath \scititle}

\author{
	Eugenio DelRe$^{1\ast}$ and
	Paolo Di Porto$^{1}$\and
	\small$^{1}$Dipartimento di Fisica, Università di Roma La Sapienza, 00185 Rome, Italy.\and
	\small$^\ast$Corresponding author. Email: eugenio.delre@uniroma1.it\and
}

\begin{document} 

\maketitle

\begin{abstract} \bfseries \boldmath
Topology is key in describing unconventional quantum phases of matter and devising robust quantum technology.  Exactly how topology mixes with quantum mechanics remains largely unclear, as testified by the lack of a unifying microscopic theory for the ever-expanding and still puzzling transport behavior of electrons in the Quantum Hall Effect.  Here we formulate a microscopic theory able to quantitatively describe the large wealth of Quantum Hall physics starting from one basic assumption, that the topological constraint in actual space leads to a superposition of states in the associated angular space. This allows us to identify the mechanism underlying quantum topology, single-particle wavefunction regularity in 3D, while  many-body physics and disorder play no fundamental role.  Our findings introduce a new  far-reaching perspective in analyzing topological quantum systems and applications, such as topological quantum computing. 
\end{abstract}

\noindent Non-trivial topology forms a powerful tool in quantum mechanics. Aside from introducing inherently new phenomena, such as the Aharonov-Bohm effect \cite{Aharonov1959}, it provides an explanatory framework for the quantization of electric charge \cite{Dirac1931,Golik2024}, magnetic flux in superconductors \cite{London1950}, and the integer quantum Hall effect (IQHE) in 2D metals and semimetals  \cite{Laughlin1981}.   This notwithstanding, the combination of topology with quantum mechanics harbors one of the principal hereto unresolved puzzles of modern physics, i.e., the fractional QHE (FQHE), with its large and ever-expanding class of unexplained results \cite{Klitzing1980,Tsui1982,Klitzing1985,Prange1990,Chakraborty1995,DasSarma1997, Stormer1999,Jacak2003,Du2009,Ghahari2011,Dean2011,Feldman2012,Willett2013,Amet2015,Dean2015,Diankov2016,Kim2019,Lu2024,Kaur2024}.     While the IQHE can be interpreted using a general single-electron argument based on gauge invariance in a two-dimensional system \cite{Laughlin1981}, the FQHE is presently believed to be connected to multi-electron strongly-correlated states \cite{Laughlin1983,Laughlin1983_1,Laughlin1998,Jain2007,Halperin2020,Feldman2021}.  This odd situation, by which the very same phenomenon appears to require two wholly contrasting models \cite{Jain1989}, one with no interaction and one with strong interaction, is further complicated by the fact that observed even-denominator fractional filling factors do not easily fit into the strongly correlated electron picture \cite{Willett1987,Jiang1989,Kang1993,Pan1999,Pan2002,Martin2004,Lin2014,Kim2019}.  Growing experimental evidence that the IQHE and FQHE share universal quantum phase-transition scaling exponents makes the idea that they arise from fundamentally different mechanisms ever more untenable\cite{Kaur2024} (see Supplementary Text section QHE: A modelling conundrum).

Apart from being an open issue, understanding how quantum mechanics affects topology, thus clarifying the principles underlying  the FQHE, is fundamental  because it forms the principal experimental playground for the growing field of topological order and topological matter, with impact on topological insulators \cite{Hasan2010,Halperin2020}, superconductors \cite{Qi2011},  and topological photonics \cite{Khanikaev2017}.  The FQHE  is also the starting point for an experimental realization of topological quantum computing,  a theoretical framework that traces a route to noise-resistant scalable quantum memory and computation \cite{Freedman2003,Bartolomei2020}.  Here,  it is exactly those little-understood even-denominator FQHE states \cite{Park2020,Pu2023} that are thought to be the prototypical embodiements of  non-Abelian anyon states, systems with long-range entanglement that are the key ingredient to topological quantum processing \cite{Nayak2008,Dutta2022}.

We here introduce the idea of quantization in topological space and discuss one of its principal consequences, that is,  that  particles confined to a 2D system are dominated by symmetry properties in 3D space.  This allows us to formulate a microscopic mechanism for the QHE whose predictions are in strong quantitative agreement with established experimental findings, including integer, odd-, and even-denominator fractional values, along with the plateau widths, of quantized Hall conductance. The resulting unified theory for the IQHE and FQHE clarifies the  connection between dimensionality, symmetry, and quantum behavior, and provides a  quantitative predictive basis for pioneering topological technology.  Specifically, quantization in topological space shows how single electrons and Cooper-pairs can behave as anyons without the introduction of strongly correlated multi-particle states, this both for odd-denominator and even-denominator Hall states, indicating a route to an accessible quantum topological computation where strongly correlated many-body systems are not required.

\noindent \textbf{Quantization in topological angular space}

\noindent Topology becomes dominant in quantum mechanics when the wavefunction of a particle is non-zero in a topologically non-trivial region of space \cite{Dirac1931, Aharonov1959}. The condition of wavefunction regularity then imposes constraints on the accumulated phase around a closed path $\Gamma$ that cannot be reduced to a point through continuous deformation, i.e., $\Delta \theta_{\Gamma}=2\pi N$, with $N$ an integer for bosons and a semi-integer for fermions \cite{Merzbacker1962,Hall2015}. Each winding along a topologically equivalent $\Gamma$ leads to the accumulated $2\pi N$, that then becomes the topological quantum.  We start from the consideration that while the topological invariant in space is $2\pi N$,  the minimum accumulated phase around $\Gamma$ that allows regularity is $\delta (\Delta \theta_\Gamma)=2\pi$.  If we assume this minimum $\delta (\Delta \theta_\Gamma)$ to be the quantum, the angular topological quantum, we find that it corresponds in actual space to a set of magnetic flux quanta that obey the relationship  

\begin{equation}
\delta \Phi(j,j_z)=(j/j_z)(hc/e),
 \label{fractionalflux}
 \end{equation}
 
\noindent  where $j$ and $j_z \le j$ are the quantum numbers of a topological total angular momentum $\mathbf{J}^2$ and its projection $J_z$, and $j\le j_z^{max}$, $j_z^{max}$ being the principal experimentally-determined parameter that depends on how strong the electrons are bound to the conducting plane (see the Quantization in topological angular space and fractional flux quanta section in Ref.{\cite{methods}} and Figs.\ref{FigureTopologicalSpace},\ref{FigureFractionalFluxQuantum}).  The result is a set of magnetic flux quanta that are, in general, larger than $hc/e$ and  fractional.

In a Hall experiment, these fractional flux quanta correspond to a fractional Hall conductance $G_{xz}=\nu(e^2/h)$, where

\begin{equation}
\nu =nj_z/j=n(q+1/2)/(p+1/2),
\label{fermions}
\end{equation}  

\noindent with $q$ and $p$ integers ($q\le p$), for fermions, and

\begin{equation}
\nu =nj_z/j=nq/p,
\label{bosons}
\end{equation}  

\noindent with $q$ and $p$ integers (and again, $q\le p$), for bosons (see the Quantization in topological space and Hall spectra section in Ref.{\cite{methods}} and Fig.\ref{Figure_Laughlin_Loop}).

In Fig.\ref{Figure_number_sequence}a we report an example of a sequence of $\nu$ with $n=1, ..., n_{max}$ and  $p=0, ..., p_{max}$ ($p_{max}+1/2=j_z^{max}$ and $p_{max}=j_z^{max}$ for the fermionic and bosonic part of the spectrum, respectively). The resulting fermionic fractional Hall conductance manifests characteristic spectrum gaps, that is, values of $\nu$ that are prohibited, and gap-center states, that is, isolated values of allowed $\nu$ that are flanked by two gaps (see the Spectrum gap and gap-center states section in Ref.{\cite{methods}}). Fermionic and bosonic $\nu$ spectra also manifest self-similarity (see the Spectrum self-similarity section in Ref.{\cite{methods}}) and a characteristic correspondence between the integer and fractional $\nu$ spectrum (see the Fractional-integer spectrum correspondence section in Ref.{\cite{methods}}). 

\noindent \textbf{Theory versus experiment}

\noindent Translating Eqs.(\ref{fermions}) and (\ref{bosons}) into predictions on measurements of $G_{xz}$ versus applied magnetic field $H_0$ and density of charge carriers $\rho$ requires the introduction of a set of experimentally determined parameters (see the Hall conductance section in Ref.{\cite{methods}} and Figs.\ref{Figure_origin_of_plateaus}, \ref{Figure_Zeeman_splitting_Graphene}). Specifically, (i) $p_{max}$, that depends on $j_z^{max}$, i.e.,  on how strong the fermionic carriers are confined to the conducting plane; (ii) the topological linewidth uncertainty $\delta \nu$, that depends on $p_{max}$ and the fermionic spectrum lineshape;  (iii) the maximum value of integer $\nu$ achievable $n_{max}$, determined by the maximum Landau-Level (LL) that can be occupied; and (iv)   the temperature-dependent ratio $\eta$ of density of electrons that condense into Cooper pairs relative to the total electron density.  For $p_{max}=0$, the involvement of spin-splitting makes predictions system-dependent with two added parameters: (v)  the value $\nu_{ss}$ at which spin-splitting intervenes and (vi) the associated reduced topological uncertainty $\delta \nu_{ss} < \delta \nu$.

\begin{figure}[t]
\centering
\includegraphics[width=0.6\columnwidth]{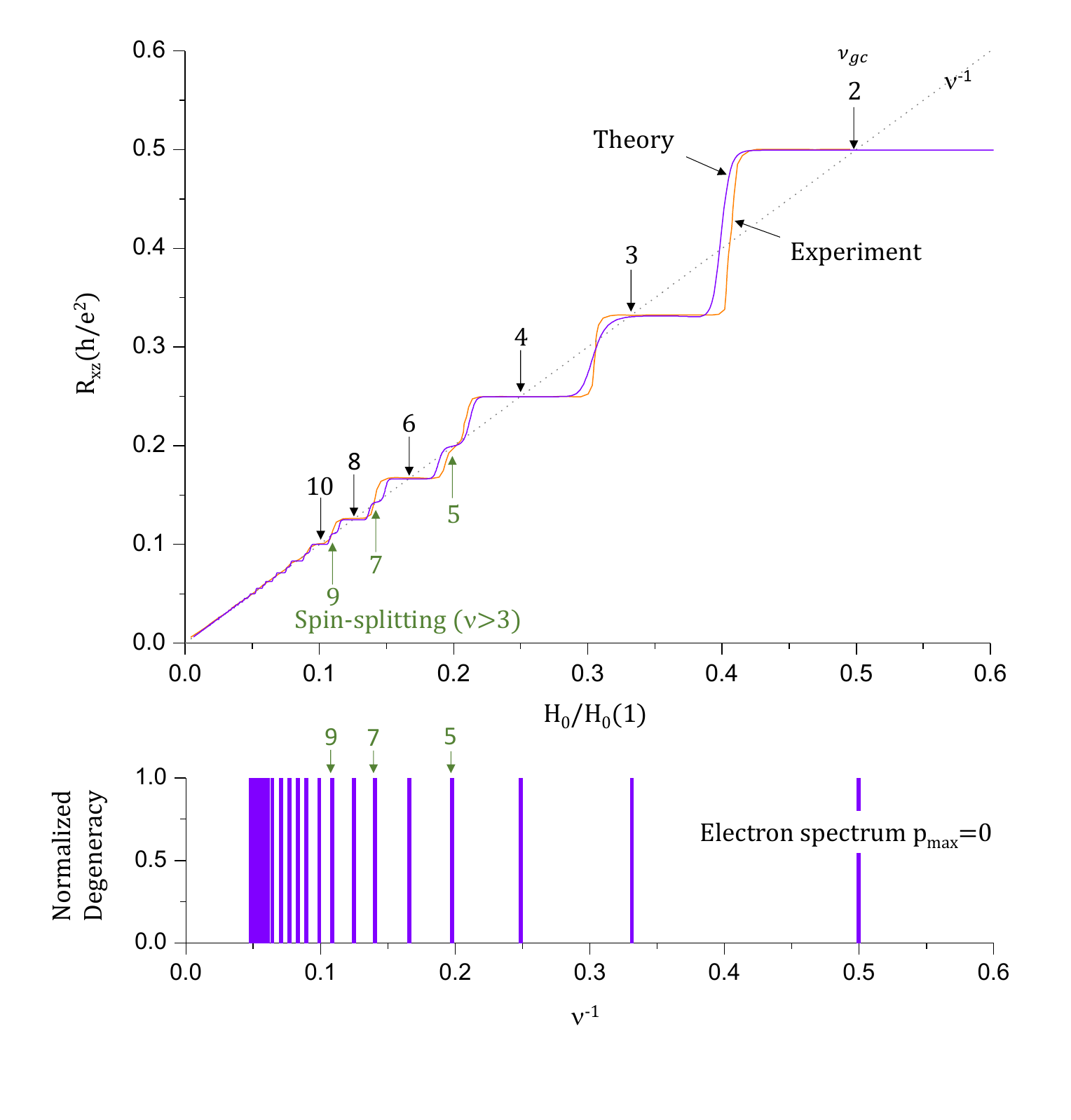}
\caption{\textbf{Theory versus experiment for the IQHE.} Prediction based on Eq.(\ref{fermions}) and corresponding Eq.(\ref{Gxz_fermionic}) pitted against measurements from Ref.\cite{Paalanen1982} (top panel, violet and orange curves, respectively).  The absence of gaps signals the spin-dominated regime of $p_{max}=0$.  Note the spin-splitting acting on the odd values for $\nu>3$ (green values). Bottom panel, the governing electron spectrum. Values of $\nu>10$ are not clear in data and are not discussed.}
\label{experiments_versus_theory_integer}
\end{figure}

We begin by testing the predictions of Eqs. (\ref{fermions}), and (\ref{bosons}), and their empirical formulation through Eqs.(\ref{Gxz_fermionic}) and (\ref{Gxz_bosonic}), with established Hall effect experimental results for $R_{xz}$ ($1/G_{xz}$) reported in  GaAs/AlGaAs heterostructures in Ref.\cite{Paalanen1982}.  Considering the experimental data reported as the orange curve in Fig.\ref{experiments_versus_theory_integer} (top panel),  it is immediately apparent that the behavior of $R_{xz}$ does not include segments corresponding to the classical Hall effect, i.e., for which $R_{xz} \propto H_0$ (see dotted $\nu^{-1}$ line, apart from the asymptotic $H_0 \rightarrow 0$ regime).  The absence of these classical-like regimes means that the system does not manifest the gap states $\nu_g$, indicating that results can either refer to a spin-dominated electron spectrum with $p_{max}=0$ or to Cooper-pairs.  While no value of $p_{max}$ makes the observed data compatible with Cooper-pair behavior (i.e., Eq.(\ref{Gxz_bosonic})), the predicted electron $R_{xz}=1/G_{xz}$ from Eq.(\ref{Gxz_fermionic}) for $p_{max}=0$, $n_{max}=160$, $\delta \nu=0.2$, $\nu_{ss}=3.5$, and $\delta \nu_{ss}=0.5$ (full violet line in Fig.\ref{experiments_versus_theory_integer} (top panel) and corresponding spectrum (bottom panel)) is in strong agreement with observations (see the Fixing $n_{max}$ section in Supplementary Text).

\begin{figure*}[t]
\centering
\includegraphics[width=1.0\columnwidth]{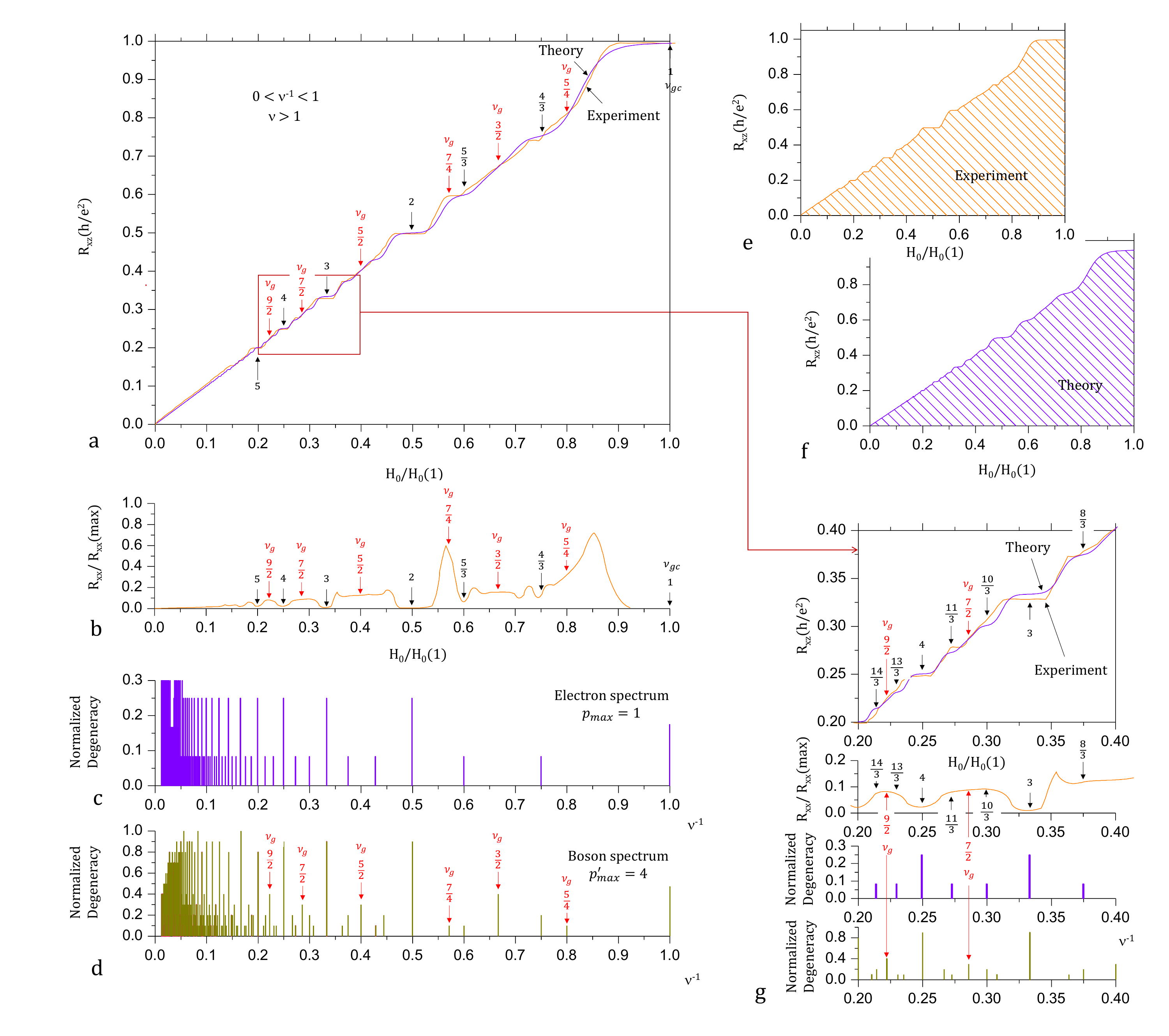}
\caption{\textbf{Theory versus experiment for the FQHE.} (a) Prediction based on Eqs.(\ref{fermions}) and (\ref{bosons}) and corresponding Eq.(\ref{Gxz_fermionic}) (violet curve) pitted against sample Hall resistance  measurements taken from Ref.\cite{Willett1987} (orange curve) for $\nu>1$ ($0<\nu^{-1}<1$). (b) Observed normalized $R_{xx}$. (c) Gap-center states $\nu_{gc}$ and gaps $\nu_g$  are referenced back to the fermionic spectrum ($p_{max}=1$) and (c) to the corresponding topologically prohibited states of the bosonic spectrum ($p'_{max}=4$).   For clarity, the comparison is halted for high values of $\nu>5$. (e), (f) Overall qualitative comparison and (g) sample expanded view of comparison down to the very minute details of $R_{xz}$.}
\label{experiments_vs_theory_fractional_first}
\end{figure*}

Next we test the predictions of our theory with the very different Hall effect results observed  in similar  GaAs/AlGaAs heterostructures in Ref.\cite{Willett1987}.  Measured values from Fig.1 of Ref.\cite{Willett1987} for cold conditions ($50$mK$<T<150$mK) are reproduced in Fig.\ref{experiments_vs_theory_fractional_first}a for $\nu>1$ (orange curve).  The $R_{xz}$ predicted on the basis of the spectrum described by Eqs.(\ref{fermions}) and (\ref{bosons}) and corresponding Eq.(\ref{Gxz_fermionic}) with $p_{max}=1$, $p'_{max}=4$,  $n_{max}=160$, and $\delta \nu=0.15$,  reported in Fig.\ref{experiments_vs_theory_fractional_first}a (violet solid line), is able to reproduce the experimental findings with remarkable accuracy.  In other words, the spectrum for single non-interacting electrons with $j_z^{max}=21/2$ is in  quantiative agreement with experiments.  Agreement is further substantiated comparing the observed normalized longitudinal resistance $R_{xx}$ (Fig.\ref{experiments_vs_theory_fractional_first}b) directly to the predicted electron and boson spectrum, reported in Fig.\ref{experiments_vs_theory_fractional_first}c,d, in the form of the normalized state degeneracy (proportional to the number of occurrences of the value in the predicted spectrum) versus $1/\nu$.   

The validity of our theory is perhaps rendered even more tangible comparing the overall features of observed and predicted $R_{xz}$.  The apparently unruly and unpredictable nature of observations (Fig.\ref{experiments_vs_theory_fractional_first}e) is matched by the once again apparently unruly behavior of predictions (Fig.\ref{experiments_vs_theory_fractional_first}f).  Figure \ref{experiments_vs_theory_fractional_first}g further reports the comparison in Fig.\ref{experiments_vs_theory_fractional_first}a, but now expanded to highlight the details in a specific segment of the scan.  Notwithstanding the increased role of limited experimental resolution, yet the agreement is still clear. 

The comparison of observations from Fig.1 of Ref.\cite{Willett1987} is reported for $2/3<\nu<1$ with $p_{max}=1$, $p'_{max}=4$,  $n_{max}=160$, and $\delta \nu=0.07$ in Fig.\ref{experiments_vs_theory_fractional_second}a-f, and  for $2/5<\nu<3/5$ with $p_{max}=5$, $p'_{max}=10$,  $n_{max}=160$, and $\delta \nu=0.005$ in Fig.\ref{experiments_vs_theory_fractional_second}g-l. The shift from $p_{max}=1$ for relatively low values of $H_0$ to $p_{max}=5$ for higher values of $H_0$ is consistent with the fact that the two experiments were carried out in the same sample but in different conditions (as described in Ref.\cite{Willett1987}).  The actual relationship between $p_{max}$ (and hence $\delta \nu$) for the different experiments remains, with the available data, to be established.

\begin{figure*}[t]
\centering
\includegraphics[width=0.8\columnwidth]{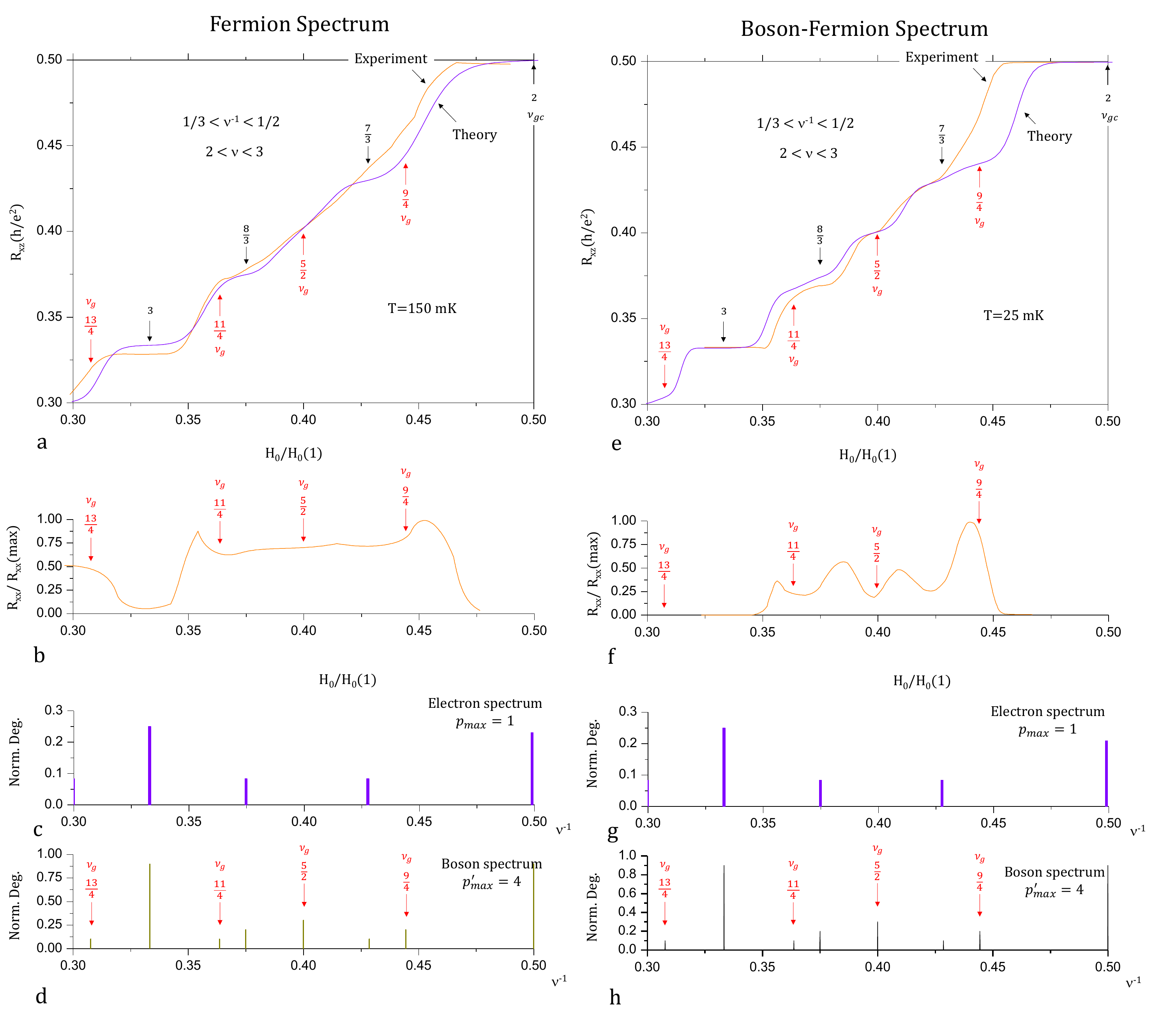}
\caption{\textbf{Evidence of bosonic spectrum.}  (a)-(d) Analysis of results at $T$=150mK  for $2<\nu<3$ from Ref.\cite{Willett1987} using Eq.(\ref{Gxz_fermionic}).  (e) Experiment (orange line)  for $T$=25mK (Fig.2 of Ref.\cite{Willett1987}) and theory (violet line) using Eq.(\ref{Gxz_bosonic}) with complete boson condensation ($\eta=0.99$).  Note the emergence of the plateau at $\nu=5/2$ at 25mK that is a topologically prohibited state at 150mK.  (f) Observed $R_{xx}$ and predicted (g) fermionic and (h) bosonic spectra.}
\label{experiments_vs_theory_boson}
\end{figure*}

The spectrum of Eqs.(\ref{fermions}) and (\ref{bosons}) in Eq.(\ref{Gxz_fermionic}) only includes potential plateaus at integer and odd-denominator fractional values of $\nu$.  Measurements of $R_{xz}$ at ultra-low temperatures ($T<50$mK) reported in Fig.2 of Ref.\cite{Willett1987} indicate that a Hall state begins to appear at $\nu=5/2$, becoming ever more evident as $T$ is decreased.  Since $\nu=5/2$ is for Eq.(\ref{fermions}) a forbidden $\nu_g$, it must mean that, as the temperature is lowered, electrons begin forming Cooper pairs with the spectrum of Eq.(\ref{bosons}) and corresponding Eq.(\ref{Gxz_bosonic}) that has the 5/2 state, i.e., that $\eta$ becomes finite. In Fig.\ref{experiments_vs_theory_boson} we compare results in the region $2<\nu<3$ for $T=150$mK (Fig.\ref{experiments_vs_theory_boson}a-d) and for $T=25$mK (Fig.\ref{experiments_vs_theory_boson}e-h).  The basic difference in experimental findings (orange full lines in Fig.\ref{experiments_vs_theory_boson}a,e) is the partial transformation of the topologically prohibited state $\nu_g=5/2$ at 150mK, with its classical like behavior, into a plateau at 25mK.   While the higher temperature data is well described by Eq.(\ref{Gxz_fermionic}) (violet curve, as in Fig.\ref{experiments_vs_theory_fractional_first}), the low temperature data is described now by both         Eq.(\ref{Gxz_fermionic}) and Eq.(\ref{Gxz_bosonic}), with the same experimental parameters found in the $\nu>1$ case of Fig.\ref{experiments_vs_theory_fractional_first}, but with a complete condensation of mobile electrons into Cooper pairs ($\eta=0.99$ for the violet full line in Fig.\ref{experiments_vs_theory_boson}e).

\begin{figure*}[t]
\centering
\includegraphics[width=0.8\columnwidth]{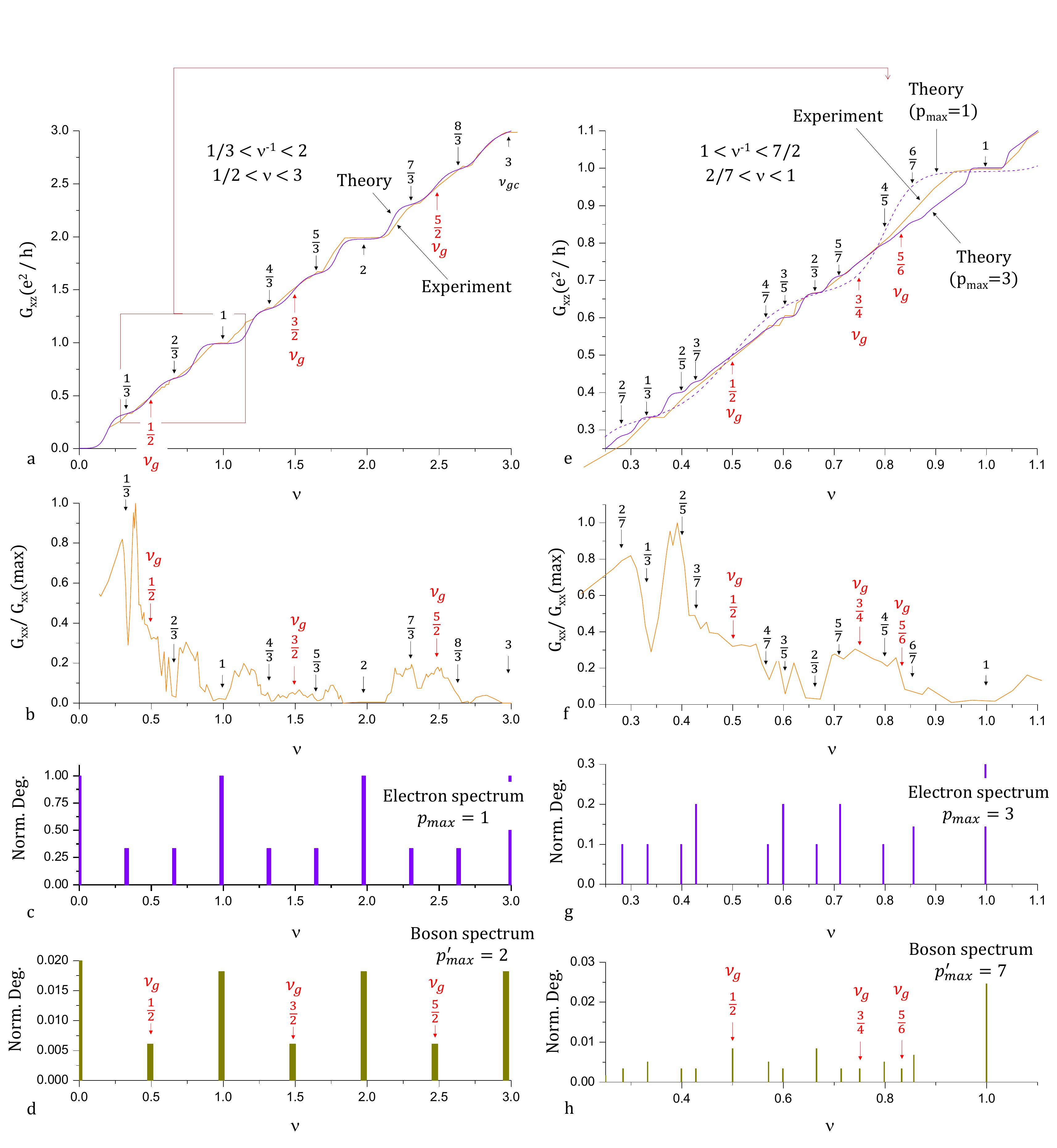}
\caption{\textbf{FQHE in Graphene.} (a) Hall conductance  measurements taken from Fig.3c of Ref.\cite{Schmitz2020} for $1/2 < \nu < 3$ (orange curve) compared to prediction based on Eqs.(\ref{fermions}) and (\ref{bosons}) and corresponding Eq.(\ref{Gxz_fermionic}) (violet curve) with $p_{max}=1$.  (b) Observed normalized $G_{xx}$. States  are referenced back to the fermionic spectrum (c) and corresponding topologically prohibited states of the (d) bosonic spectrum ($p'_{max}=2$). (e)-(h) Analogous comparison for the extreme quantum regime $2/7<\nu<1$ indicating a higher $p_{max}=3$ (see text).}
\label{experiments_vs_theory_graphene}
\end{figure*}

We next test our theory against the Hall effect observed in graphene. We begin by discussing how quantization in topological space, with its associated topological uncertainty, allows the interpretation of the so-called anomalous QHE in SLG and unconventional QHE in BLG (see the Supplementary Text section Anomalous and Unconventional QHE). A more detailed and hence stonger quantitative test of our theory is achieved analyzing the FQHE in graphene \cite{Du2009,Dean2011,Ghahari2011,Amet2015,Diankov2016,Kim2019,Schmitz2020,Lu2024}.    We focus on the highly reproducible and sample-independent results found in CVD-grown SLG reported, for example, in Fig.3c of Ref.\cite{Schmitz2020}.  
In Fig.\ref{experiments_vs_theory_graphene}a we compare observed $G_{xz}$ versus $\nu$ to the model prediction with $p_{max}=1$, $p'_{max}=2$,  $n_{max}=80$, and $\delta \nu=0.15$. The agreement is, once again, clear.  The high resolution of reported results also in the $G_{xx}$ versus $\nu$ (Fig.\ref{experiments_vs_theory_graphene}b) allows us to push our analysis further.  Specifically, in the extreme quantum regime of lower values of $\nu <1$, that would correspond to truly elevated values of $H_0$ in a standard $H_0$ scan, there is an emerging spectral structure that violates the predicted $p_{max}=1$, $p'_{max}=2$ spectrum (Fig.\ref{experiments_vs_theory_graphene}c,d).  Apart from the minor detail of the width of the plateau at $\nu=1$, two distinct plateaus in $G_{xz}$ with their corresponding dips in $G_{xx}$ are evident in the region $1/2<\nu<2/3$.  As for the above discussed semiconductor heretrostructures, so in the present case our theory indicates that the passage into the extreme quantum regime involves an increase in $p_{max}$.  The boxed out region in Fig.\ref{experiments_vs_theory_graphene}a  is analyzed in detail in Fig.\ref{experiments_vs_theory_graphene}e-h, and is evidently compatible with a higher $p_{max}=3$, $p'_{max}=7$, the new spectral features coinciding then with the $\nu=4/7$ and $\nu=3/5$ states.  Importantly, the $p_{max}=3$ spectrum should lead to added plateaus at $\nu=4/5$ and $\nu=6/7$, that are not observed, plateaus that are obfuscated in the model by the topologically prohibited states at $\nu=3/4$ and $\nu=5/6$, a further detailed confirmation of the reasoning leading to Eq.(\ref{Gxz_fermionic}). In turn, predictions indicate that observable plateaus should also form at $\nu=2/5$ and $\nu=3/7$ that are not, as yet, observed.  This picture, i.e., an overall agreement between experiments and theory (the $p_{max}=1$ model), its partial falsification (in the extreme $\nu<1$ region), its consequent modification (the $p_{max}=3$ model), and, finally, its use as a predictive instrument, fits into the basic paradigm of what we would consider a useful physical theory.  

The fact that our theory finds agreement also in graphene has one immediate unexpected consequence:  in analogy to the reasoning leading to the emergence of a bosonic spectrum at ultra-low temperatures in semiconductor heterostructures (Fig.\ref{experiments_vs_theory_boson}), evidence of even denominator states in graphene at very low temperatures \cite{Zibrov2018,Kim2019} requires both         Eq.(\ref{Gxz_fermionic}) and Eq.(\ref{Gxz_bosonic}), i.e., a finite value of $\eta$.  The implication is that our theory predicts that graphene, in these conditions, should support something equivalent to the formation of Cooper pairs.

 \noindent \textbf{Discussion}

\noindent Strong agreement of our theory with both IQHE and FQHE experiments indicates that these can arise from a single physical mechanism, quantization in topological angular space, and involve non-interacting electrons and Cooper pairs.  The IQHE theory developed by Laughlin with the introduction of a single-loop geometry (see $\Gamma$ in Fig.\ref{Figure_Laughlin_Loop}) \cite{Laughlin1981} amounts to considering solely $j_z=j$ in Eq.(\ref{fundamental}).  This is a correct assumption if the space hosting the loop is in fact 2D, in which case $J$ and $J_z$ coincide.  Conditions in which $j_z \neq j$, that lead to fractional  values of the flux quantum, are then the signature of the fact that the space hosting the path $\Gamma$ is intrinsically 3D. Moreover, our theory describes well the emergence of even-denominator fractional values of $\nu$, attributing these to transport by bosonic quasi-particles, overcoming a basic challenge of previous approaches \cite{Willett1987,Jiang1989,Kang1993,Pan1999,Pan2002,Martin2004,Lin2014}.   As regards to the Hall plateau distribution and width, while present models attribute their formation  to a not well-specified disorder, our theory is able to predict the details of $G_{xz}$, i.e., which plateaus form and which do not and why, which regions lead to classical-like behavior and which do not and why, and this across different materials and experiments with one principal experimentally established parameter, $p_{max}$, physically connected to how strong the transport particles are bound to the conducting plane.  Further, our theory naturally explains, as a direct consequence of quantization in topological space, spectrum self-similarity, previously attributed to electron-electron interaction \cite{Mani1996}, and the IQHE-FQHE spectrum correspondence, an unexplained feature taken as the phenomenological basis for the so-called composite-Fermion theory. Finally, our theory explains QHE observed in graphene, including anomalous, unconventional, and fractional Hall conductance, unifying it with effects observed in semiconductor heterostructures. 

Previous attempts to explain fractional behavior introduce interaction as an added ingredient: for quantization in topological angular space, the added ingredient is to consider the full 3D nature of symmetry even for particles constrained to a plane.  In this, our theory reinstates the basic quantum-mechanical  notion that electrons confined to a plane are still in 3D space, as required by the uncertainty principle. Considering then for electrons constrained to a plane only a 2D  space implies capturing only part of the behavior, i.e., the IQHE, while extraneous mechanisms, such as many body correlation, are required to reproduce that part of behavior, i.e., the FQHE, that stems from the full 3D nature of the system. 

The result is a fundamentally altered picture of topologically dominated quantum electron transport.  For one,  non-Abelian anyons, key to the conceptual development of topological quantum computing, do not require many-body physics (see the Non-interacting particles and anyons section in Supplementary Text).

\clearpage 

%

%
%


\section*{Acknowledgments}

\paragraph*{Funding:}
 E.D.R. acknowledge support from the PNRR  MUR Project No. PE0000023-NQSTI, the PRIN2022 MUR Project No. 20223T577Z, and Sapienza- Ricerca di Ateneo Projects.
\paragraph*{Author contributions:}
Authors contributed equally.
\paragraph*{Competing interests:}
There are no competing interests to declare.


\subsection*{Supplementary materials}
Materials and Methods\\
Supplementary Text\\
Figs. S1 to S7\\
References \textit{(7-\arabic{enumiv})}\\ 


\newpage


\renewcommand{\thefigure}{S\arabic{figure}}
\renewcommand{\thetable}{S\arabic{table}}
\renewcommand{\theequation}{S\arabic{equation}}
\renewcommand{\thepage}{S\arabic{page}}
\setcounter{figure}{0}
\setcounter{table}{0}
\setcounter{equation}{0}
\setcounter{page}{1} 


\begin{center}
\section*{Supplementary Materials for\\ \scititle}

E. DelRe$^{\ast}$,
P. Di Porto\\ 
\small$^\ast$Corresponding author. Email: eugenio.delre@uniroma1.it\\
\end{center}

\subsubsection*{This PDF file includes:}
Materials and Methods\\
Supplementary Text\\
Figures S1 to S7\\

\newpage


\subsection*{Materials and Methods}

\subsubsection*{Quantization in topological angular space and fractional flux quanta}

\begin{figure}[t]
\centering
\includegraphics[width=1.0\columnwidth]{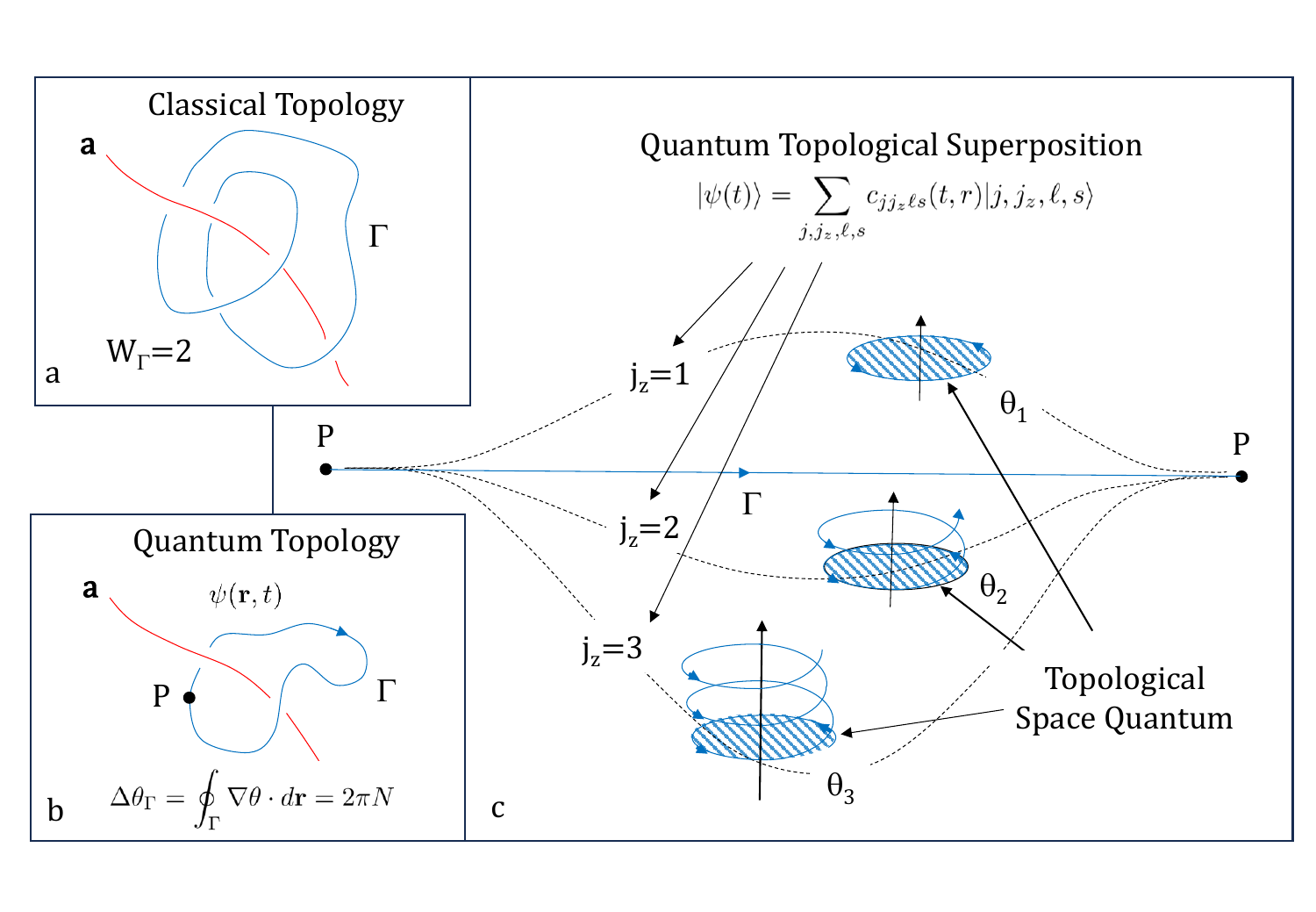}
\caption{\textbf{Quantization in topological space.} (a) Classical Topology and the invariant number of times $W_\Gamma$ the curve $\Gamma$ winds around the forbidden infinite curve $a$.  (b) Single and double-valuedness in the wavefunction $\psi$ introduces a quantum topological invariant in the form of the accumulated phase $\Delta \theta_\Gamma$ along a closed path $\Gamma$. (c) Path $\Gamma$, represented by the segment in actual space from $P$ to $P' \equiv P$, is characterized by the superposition of different associated paths in corresponding angular spaces $\theta_1$, $\theta_2$, and $\theta_3$ for the case of $j=3$. Shaded  regions represent the three different angular topological quanta $\delta (\Delta \theta_\Gamma) (j=3,j_z=1)=2\pi (3/1)$, the fractional $\delta (\Delta \theta_\Gamma)(3,2)=2\pi (3/2)$, and $\delta (\Delta \theta_\Gamma)(3,3)=2\pi (3/3)$ (see Eq.(\ref{anglequanta})).}
\label{FigureTopologicalSpace}
\end{figure}

\noindent \textit{Quantization in topological angular space}. We recall that a topological quantity is invariant with respect to a set of continuous deformations. An  example is  a closed curve $\Gamma$ around an infinite (or closed) forbidden line $a$.  The number $W_\Gamma$ of times $\Gamma$ winds around $a$ is topological, in that it does not change however we continuously stretch or squeeze $\Gamma$ (see illustration in Fig.\ref{FigureTopologicalSpace}a).  

Consider the quantum topological property that arises from the continuity conditions that the wavefunction $\psi(\mathbf{r},t)$, representing the state of a system, must satisfy \cite{Dirac1931,Aharonov1959,Merzbacker1962}.  For the case of a particle without spin,  $\psi (\mathbf{r},t)=|\psi (\mathbf{r},t)|\exp{(i\theta(\mathbf{r},t))}$ must be single-valued also in the presence of a forbidden line $a$, i.e., one for which $\psi(\mathbf{r},t)=0$ $\forall \mathbf{r} \in a$.  The condition implies that the phase accumulated by the wavefunction along a closed path $\Gamma$ that loops around $a$ is  

\begin{equation}
\Delta \theta_\Gamma=\oint_\Gamma \nabla \theta \cdot d\mathbf{r}=2\pi N,
\label{phase}
\end{equation}

\noindent with $N$ an integer. This suggests that $N$ is topological, in that deforming $\Gamma$, without passing through $a$, cannot change its value (see Fig.\ref{FigureTopologicalSpace}b).   For a spin $1/2$ particle with  $|\psi\rangle=(\psi^{(1)}(\mathbf{r},t),\psi^{(2)}(\mathbf{r},t))$, where $\psi^{(1)}$ and $\psi^{(2)}$ are the amplitudes for the spin up and spin down cases in the Pauli matrix representation, the phase accumulated along the path $\Gamma$ is such as to change the sign of each component (i.e., $|\psi\rangle \rightarrow -|\psi\rangle$), so that, for each component,  $\Delta \theta_\Gamma=\oint_\Gamma \nabla \theta \cdot d\mathbf{r}=2\pi N$, with $N$ now a half-integer.  The double-valuedness of each single component, in this case, is the direct result of the fact that moving along $\Gamma$ involves a rotation in physical space that, for the spinor, leads to a double-valued projective representation in 3D space \cite{Hall2015}.
Analogously, $N$ is once again an integer for the three components of a spin 1 particle.

To introduce the idea of quantization in topological space, we begin by noting that, in terms of topology in actual space,  the condition of Eq.(\ref{phase}) implies a minimum topological quantum of accumulated phase $\Delta \theta_\Gamma=2\pi N$, since each winding around $a$ corresponds to one full winding along $\Gamma$.  In turn,  wavefunction regularity will hold for any integer (or half-integer) value of $N$. It follows that the spectrum of allowed states, for both integer and half-integer spin systems, spans all values of $N$, each state separated by the smaller quantum in accumulated phase $\delta (\Delta \theta_\Gamma) =2\pi$ that corresponds to the passage from a given $N$ to $N\pm 1$.  This  quantum is evidently not the topological quantum in actual space. What is it the topological quantum of? Since $\delta (\Delta \theta_\Gamma)=2\pi$ naturally forms the topological quantum in the angular space in which $\theta$ itself winds as the path $\Gamma$ winds in actual space, an intuitive hypothesis is then that the topological quantum refers to regularization in a closed path in the angle space of $\theta$.    To determine the nature of this angle space and its relationship to actual space,   we note that the most general expression for an arbitrary state obeying the single- or double-valuedness of Eq.(\ref{phase}) is 

\begin{equation}
|\psi(t)\rangle = \underset{j,j_z,\ell,s}{\sum}c_{jj_z\ell s}(t,r)|j,j_z,\ell,s\rangle . 
\label{superposition}
\end{equation}

\noindent Here we use the spherical coordinates $r$, $\vartheta$, and $\varphi$  with origin in a specific point $P$ of the curve $a$, say at $\mathbf{r}_0$, and $|j,j_z,\ell,s\rangle$ are the eigenstates of the ordinary total angular momentum $\mathbf{J}=\mathbf{L}+\mathbf{S}$ (orbital and spin, in units of $\hbar$) relative to $\mathbf{r}_0$ with z-component $J_z$, where $\mathbf{u}_z$ is the unit vector tangent to $a$ in $\mathbf{r}_0$. The indices $j,j_z,\ell,s$ correspond to the complete set of observables $\mathbf{J}^2$, $J_z$, $\mathbf{L}^2$, $\mathbf{S}^2$, with eigenvalues $j(j+1)$, $j_z$, $\ell(\ell+1)$, and $s(s+1)$, respectively. Each term in the sum of Eq.(\ref{superposition}), being an eigenstate of $\mathbf{J}^2$ and $J_z$, has a $\varphi$ dependence $\exp{(ij_z\varphi)}=\exp{(i(m+k)\varphi)}$, where $m$ is the eigenvalue of $L_z$ and $k$ of $S_z$, while $-j\le j_z \le j$, and $|\ell-s|\le j \le \ell + s$.  This means that the accumulated phase along $\Gamma$ associated to $\varphi$ of each term is $2\pi j_z=2\pi(m+k)$. If $\Gamma$ is  a linked circumference of radius $R$ centered in $P$ at $\mathbf{r}_0$ on the plane normal to $\mathbf{u}_z$,  the total phase accumulated along $\Gamma$ by each term is exactly this $2\pi (m+k)$.  If $\Gamma$ is now deformed without crossing $a$, this accumulated phase cannot change. It follows that for a generic $\Gamma$ wrapped around $a$, each term in Eq.(\ref{superposition})  is characterized by the topological invariant  $j_z=m+k$. In terms of the spectrum of allowed states, each eigenstate of $J_z$ is characterized by a different correspondence between actual space and angular space, as $j_z$ determines  the number of times $\theta_{m+k}=(m+k)\varphi$ winds around its origin as the path $\Gamma$ winds around $a$.  Wavefunction regularity then translates into the quantum superposition of Eq.(\ref{superposition}) of different angular winding paths, each path  corresponding to an eigenstate of the relevant topological quantity  $J_z$. Since $J_z$ is also the total angular momentum component along $\mathbf{u}_z$, it has all its properties, spectrum, and uncertainty relations. In particular, its uncertainty derives  from that of ordinary angular momentum:  if $|\psi(t)\rangle$ describes a system bound to a plane that contains $\mathbf{r}_0$ and is normal to $\mathbf{u}_z$ (say, $\pi(\mathbf{u}_z)$), the spectrum of $J_z$ will be limited to  $-j_z^{max} \le j_z\le +j_z^{max}$,  where $j^{max}_{z}$ is connected to the uncertainty principle through the inequality $ (1/2) \langle J_{z} \rangle  \le \sigma_{J_x}\sigma_{J_y}$ \cite{Cohen-Tannoudji1977}, where the values of the uncertainties $\sigma_{J_x}$ and $\sigma_{J_y}$ are determined by how strong the particle described by $|\psi \rangle$ is confined to $\pi(\mathbf{u}_z)$. It follows that for a wavefunction confined to a plane, the spectrum will be characterized by all values of $j_z$ associated to values of $j$ such that $j\le j_z^{max}$.

Now, the accumulated phase along $\Gamma$ of Eq.(\ref{phase}) maps 

\begin{equation}
 |\psi\rangle \rightarrow \exp{(i\Delta \theta_\Gamma)}|\psi\rangle 
\label{mapping1}
\end{equation}
\noindent that, using Eq.(\ref{superposition}), means
\begin{equation}
 |\psi\rangle \rightarrow \underset{j,j_z,\ell,s}{\sum}c_{jj_z\ell s}(t,r)\exp{(i\Delta \theta_\Gamma)}|j,j_z,\ell,s\rangle.
\label{mapping2}
\end{equation}
\noindent Since each component of Eq.(\ref{superposition}) is mapped $|j,j_z,\ell,s\rangle \rightarrow \exp{(ij_z\delta  (\Delta  \theta_\Gamma)(j,j_z))} |j,j_z,\ell,s\rangle$, the mapping is also
\begin{equation}
 |\psi\rangle \rightarrow  \underset{j,j_z,\ell,s}{\sum}c_{jj_z\ell s}(t,r)\exp{(ij_z\delta (\Delta \theta_\Gamma)(j,j_z))}|j,j_z,\ell,s\rangle,
\label{mapping3}
\end{equation}

\noindent meaning that  $j_z\delta (\Delta \theta_\Gamma)(j,j_z)=j\delta (\Delta \theta_\Gamma)(j,j_z=j)=\Delta \theta_\Gamma$, for any $j_z \le j$.  It follows that $\delta (\Delta \theta_\Gamma)(j,j_z=j)$ must correspond to the minimum quantum of accumulated phase $2\pi$ for both bosons and fermions, i.e., the minimum passage in accumulated phase from $N$ to $N\pm 1$ in Eq.(\ref{phase}), and 

\begin{equation}
\delta (\Delta \theta_\Gamma)(j,j_z)=2\pi (j/j_z).
\label{anglequanta}
\end{equation}

\noindent Hence, the hypothesis of a topological quantum in angular space implies that the spectrum of acceptable states will have an entire family of topological quanta in actual space, only the one for $j=j_z$ resulting in a $\delta (\Delta \theta_\Gamma)(j,j)=2\pi$.  For example, a $2\pi$ quantum in the path associated to $j_z=3/2$ and $j=5/2$ corresponds to a $\delta (\Delta \theta_\Gamma)(j=5/2,j_z=3/2)=(2\pi)(5/3)$, while only the quantum associated to the path of $j_z=5/2$ and $j=5/2$ corresponds to $\delta (\Delta \theta_\Gamma)(5/2,5/2)=2\pi$.  That $\delta (\Delta \theta_\Gamma)(j=5/2,j_z=3/2)$ is fractional does not, in turn, violate the regularity in actual space of Eq.(\ref{phase}), since the accumulated phase along $\Gamma$ in actual space is $j_z\delta (\Delta \theta_\Gamma)(j=5/2,j_z=3/2)=(3/2)(2\pi)(5/3)=(5\pi)$, as expected for a fermion.  Analogously,  a quantum in the path associated to a $j_z=2$ and $j=3$ corresponds to a $\delta (\Delta \theta_\Gamma)(j=3,j_z=2)=(2\pi)(3/2)$, while only the quantum associated to the path of $j_z=3$ and $j=3$ corresponds to $\delta (\Delta \theta_\Gamma)(3,3)=2\pi$.  The fractional $\delta (\Delta \theta_\Gamma)(j=3,j_z=2)$ leads to an accumulated phase along $\Gamma$ of $j_z\delta (\Delta \theta_\Gamma)(j=3,j_z=2)=2(2\pi)(3/2)=(2\pi)3$, in agreement with the regularity conditions for a boson (see Fig.\ref{FigureTopologicalSpace}c).

In summary, the 3D nature of physical space implies that the associated quantum topological angular space is also 3D. Since this space refers to phase accumulated along circular trajectories, it is described through the formalism of 3D total angular momentum. This, as described in the following, allows us to interpret the vast and puzzling experimental results on the FQHE without introducing interactions between charged particles.

\noindent \textit{Fractional flux quanta}. Consider the situation in which $|\psi \rangle$ represents a particle of charge $Q$ and mass $M$ in a magnetic field $\mathbf{H}$.  If $\mathbf{H}$ is confined to the prohibited region at $\mathbf{r}_0$, the accumulated phase around the closed path $\Gamma$  can also be associated to the flux through the path, $\Phi_\Gamma$, by the relationship

\begin{equation}
\Delta \theta_\Gamma=(Q/\hbar c)\oint_\Gamma \mathbf{A} \cdot d\mathbf{r}=(Q/\hbar c )\Phi_\Gamma,
\label{AB}
\end{equation}
where  $\mathbf{H}=\nabla \times \mathbf{A}$ and $\mathbf{A}$ is the vector potential  \cite{Aharonov1959,Merzbacker1962}. As for Eq.(\ref{anglequanta}), but now in terms of flux, given the superposition (Eq.(\ref{superposition})), all $j_z$-components share the same experimentally determined flux $\Phi_\Gamma$ for all possible values of $j_z$.   It follows that $j_z\delta \Phi (j,j_z) =j \delta \Phi (j,j)$,  which implies that $\delta \Phi (j,j)$ must be the smallest of all flux quanta.  Its value is then determined by the minimum flux quantum compatible with $2\pi=(Q/\hbar c)\delta \Phi (j,j)$, i.e., $\delta \Phi (j,j)=hc/Q$ and  $\delta \Phi (j,j_z)=(j/j_z)(hc/Q)$.  For $Q=-e$, the spectrum of permissible states of Eq.(\ref{anglequanta}) then becomes, in terms of flux quanta, Eq.(\ref{fractionalflux}).

\begin{figure}[t]
\centering
\includegraphics[width=1.0\columnwidth]{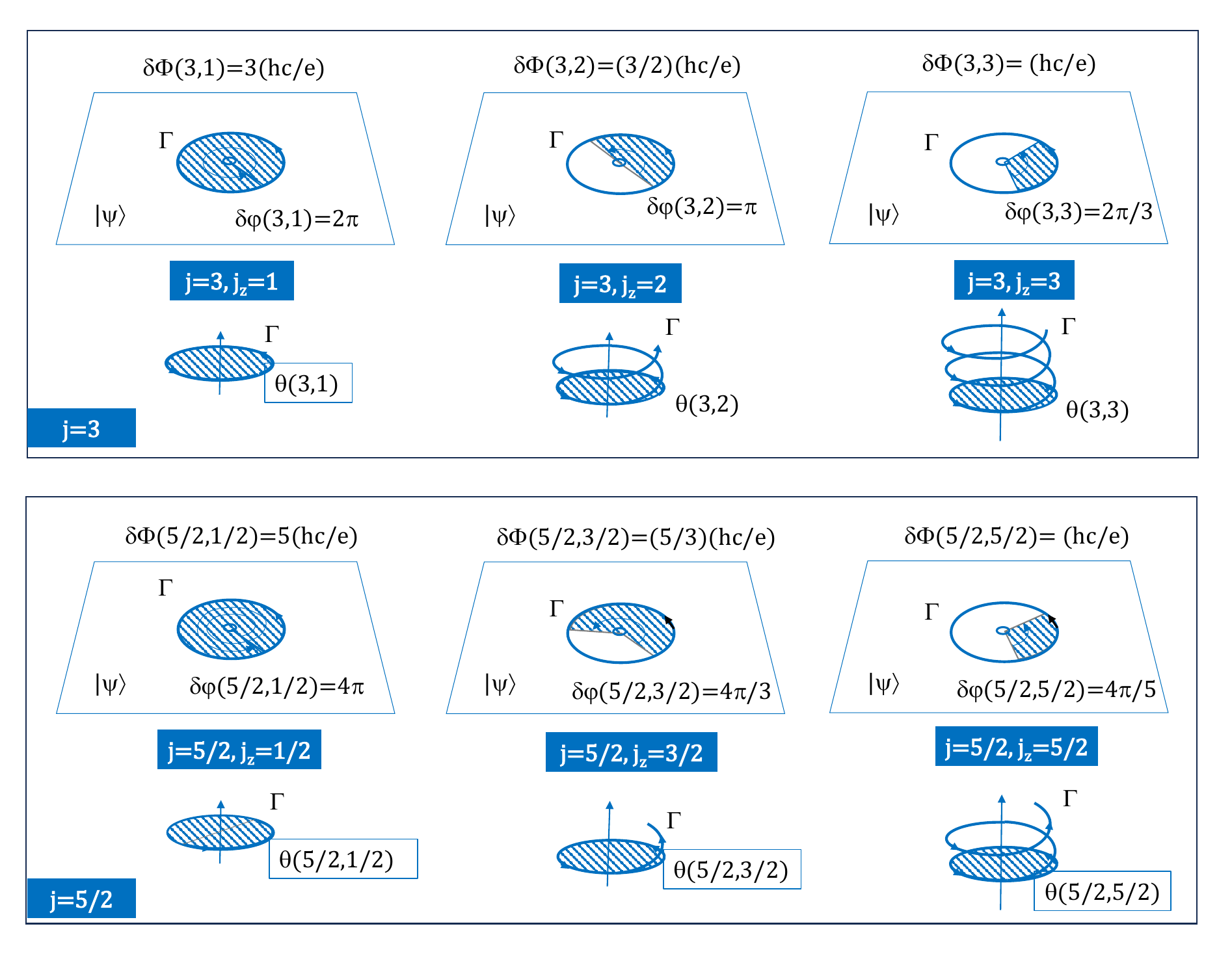}
\caption{\textbf{Fractional flux quantum for a charged particle in a magnetic field.} (First row) Illustration of a boson case for $j=3$. For $j_z=1$, path $\Gamma$ in actual space (top-left) also corresponds to a full turn in associated angle space for $\theta$ (bottom-left) and the flux quantum, associated to a full turn $\delta \theta (j,j_z)=2\pi$ (shading bottom panel), corresponds to $\Gamma$ itself in actual space, since $\delta \varphi (j,j_z)=2\pi/j_z$ is $\delta \varphi (3,1)=2\pi$ (shading top panel), and $\delta \Phi (3,1)=3(hc/e)$. For $j_z=2$ and $j_z=3$, the flux quantum now corresponds to $\delta \varphi (3,2)=2\pi/2=\pi$ and $\delta \varphi (3,3)=2\pi/3$, a slice that is a 1/2 and a 1/3 fraction of the full turn in actual space, respectively. (Second row) Illustration of a fermion case for $j=5/2$.  Note the $j_z=1/2$ case, in which the trajectory $\Gamma$ in actual space only corresponds to half a turn in angle space (bottom-left), so that the flux quantum corresponds to a double $\Gamma$ trajectory in actual space, $\delta \varphi (5/2,1/2)=4\pi$ (top-left).}
\label{FigureFractionalFluxQuantum}
\end{figure}

As for the previously discussed fractional topological quanta of Eq.(\ref{anglequanta}), the counterintuitive fractional $\delta \Phi (j,j_z)$, for $j_z<j$, can be understood considering  that each $j_z$-component has its topological quantum determined by a full $2\pi$ turn in its associated $\theta_{j_z}$ angular space, a quantum that is in no way fractional.  What is fractional is the corresponding  flux quantum as determined through Eq.(\ref{AB}).  Consider, for simplicity, a circular $\Gamma$:  the quantum $\delta \Phi (j,j_z)$  associated to a $2\pi$ full turn in $\theta_{j_z}$  corresponds to moving along an arc of $\Gamma$, starting for example at $\varphi=0$ and ending at $\varphi=2\pi/j_z$ ($\theta_{j_z}=j_z\varphi$). In other words, a quantum of flux that corresponds to a full turn in the topological angular space representing $\theta_{j_z}$ translates to a flux through a slice of the region contained by $\Gamma$, i.e., not simply a full turn in actual space  (see Fig.\ref{FigureFractionalFluxQuantum}).  Another way of seeing this is  that each $j_z$-component is periodic, so that the topological single-valuedness condition that leads to the flux quantum must be imposed not on  $\Gamma$, but on the finite arc with $\delta \varphi=2\pi/j_z$.  Since the flux quantum $hc/e$ corresponds to the slice-quantum for the $j$-component with $\delta \varphi =2\pi/j$, the flux quantum for the $j_z$-th component with $\delta \varphi =2\pi/j_z$ must lead, in real space, to a flux quantum that is $(j/j_z)$ that of the one corresponding to the  $2\pi/j$ slice-flux, i.e., Eq.(\ref{fractionalflux}). Evidently, if $j=j_z$, as forcibly occurs for an actual space that is 2D, only the well-known flux quantum $hc/e$ emerges from Eq.(\ref{fractionalflux}), in agreement with previous theories of the IQHE \cite{Laughlin1981}.

\subsubsection*{Quantization in topological space and Hall spectra}

As illustrated in Fig.\ref{Figure_Laughlin_Loop}a, in a Hall experiment, an electric field causes electrons bound to a planar metallic stripe (shaded) to flow with a current intensity $I$ along the $\mathbf{u}_x$ direction of length $\ell$ in the metal.  An intense constant magnetic field $\mathbf{H}_0$ is directed normal to the stripe, the result being a detectable Hall potential difference $V$ measured across the stripe along $\mathbf{u}_z$, the direction of the electromotive Hall force $\mathbf{F}=(Q/c)\mathbf{v}\wedge \mathbf{H}_0$, where $\mathbf{v}$ is the drift velocity and $Q=-e$ is the electron charge. The vectors $Q\mathbf{v}$ and $\mathbf{H}_0$ identify a plane, normal to $\mathbf{u}_z$.  Measurements involve the so-called  Hall conductance $G_{xz}=I/V$ in various conditions of electron temperature, stripe thickness, carrier density, and magnetic field intensity. At low temperatures and high magnetic fields, the Hall conductance shows strikingly precise plateaus versus magnetic field intensity at values of $G_{xz}=I/V=\nu e^2/h$, where $\nu$ is the so-called  integer or fractional filling factor. 

As originally noted by Laughlin, the Hall experiment of Fig.\ref{Figure_Laughlin_Loop}a becomes an experiment on quantum topology as soon as the  temperature is low enough so that the conduction electron wavefunction $\rvert \psi \rangle$ extends throughout the entire circuit \cite{Laughlin1981}.  The circuit then becomes physically equivalent to a loop path $\Gamma$  as illustrated  in Fig.\ref{Figure_Laughlin_Loop}b (Laughlin loop geometry).  As illustrated in Fig.\ref{Figure_Laughlin_Loop}c, the magnetic field through the stripe $\mathbf{H}_0$ becomes linked to $\Gamma$, with the linked magnetic flux $\Phi_\Gamma$ equal to the original flux through the sample itself. In these terms, the  current around the path, $I$, is  connected to the change in flux through the loop $\delta \Phi$ by Faraday's Law $I=c\delta U/\delta \Phi$, where $\delta U$ is the variation in energy of the system.  Since the family of allowed values of linked magnetic field compatible with delocalized electron wavefunctions must obey regularity,  values of $\delta \Phi$ that lead to observable effects will obey Eq.(\ref{fractionalflux}).  Recasting the Hall geometry in terms of the planar path $\Gamma$ around the forbidden point $P$ (see Fig.\ref{Figure_Laughlin_Loop}d), insomuch that changes in flux  act in a manner analogous to the Aharonov-Bohm effect for the path $\Gamma$,  each changing magnetic flux quantum is formally equivalent to a gauge transformation that remaps the ribbon system into itself. This means that $\delta U$ can only be associated to the passage of an integer number of electrons $n$ from one edge of the ribbon to the other along  $\mathbf{u}_z$.  Hence $\delta U=neV$, so that 

\begin{equation}
G_{xz}=I/V=cne/\delta \Phi.
\label{HallConductance}
\end{equation}
Substituting  into Eq.(\ref{HallConductance}) the fractional flux quantum of Eq.(\ref{fractionalflux}) then gives

\begin{equation}
G_{xz}=I/V=cne/\delta \Phi_{j_z}=n(e^2/h)(j_z/j)=\nu (e^2/h),
\label{fundamental}
\end{equation}   

\noindent where $\nu =n j_z/j$  and $n$ is an integer. The set of filling factors that support extended states can be further classified as fermionic and bosonic.  For half-integer values of $j$, the filling factor in Eq.(\ref{fundamental})  becomes  Eq.(\ref{fermions}), while for integer values of $j$, it becomes Eq.(\ref{bosons}).

\subsubsection*{Spectrum gap and gap-center states}

A first important feature is the appearance of recurrent gap states in the spectrum, that is, regions for which no extended quantum Hall states form in the fermionic part of the spectrum, this for an arbitrary $n_{max}$ and $p_{max}$.  A paradigmatic example at $\nu_g=1/2$ is reported in Fig.\ref{Figure_number_sequence}b.  The gaps in the fermionic spectrum are the result of a peculiar property of Eq.(\ref{fermions}), that is,  that  $\nu \neq \nu_g \equiv (2s+1)/(2r)$ with $s$ and $r$ integers.  This is because $\nu = (2s+1)/(2r)$ would imply, through Eq.(\ref{fermions}),   $2nr(2q+1)=(2s+1)(2p+1)$, that is manifestly impossible.  Hence, for example, no fermionic Hall states  are expected in and around $\nu_g =1/2$ ($s=0$, $r=1$), $\nu_g=1/4$ ($s=0$, $r=2$), $\nu_g=3/4$ ($s=1$, $r=2$).   The width of the gaps depends specifically on $p_{max}$ (i.e., on $j_z^{max}$).  To illustrate this, consider that Eq.(\ref{fermions}) implies that the spectrum is explored in steps of $1/(2p+1)$, with $p\le p_{max}$.  Since the spectrum necessarily includes the integer values $\nu=m$ as long as $n_{max}\ge m$, as follows from the fact that $n(2q+1)=m(2p+1)$ is always satisfied for $n=m$,  each of these sequences will employ an odd number of steps to go from $\nu=m-1$ to $\nu=m$ ($m\ge 1$). It follows that half-integer values of the $\nu_g$ set, such as 1/2, 3/2, 5/2, and so on, lie exactly at a half-point of the minimum step $\Delta \nu=1/(2p_{max}+1)$, meaning that the gap itself is of width $\Delta \nu$.  Congruently, in the example of Fig.\ref{Figure_number_sequence}b (top), $p_{max}=20$ and the gap at $\nu_g=1/2$ is $\Delta \nu =1/(2p_{max}+1)=1/41 \simeq 0.024$ wide.  The gap states, in turn, are not present in the corresponding bosonic part of the spectrum, in agreement with the fact that in Eq.(\ref{bosons}) $p$ and $q$ have no definite parity.  More precisely, the gap states in the fermionic spectrum coincide, in the bosonic spectrum, with the appearance of a quite remarkable spectral structure: a gap with a single isolated gap-center state, the isolated state positioned at the center of its gap, as shown in Fig.\ref{Figure_number_sequence}b (bottom).  Analogous gap-center states also form a recurrent feature in the fermionic part of the spectrum.  Tagging the fermionic gap-center states $\nu_{gc}$, one example is reported in Fig.\ref{Figure_number_sequence}c (top), at $\nu_{gc} =1$. The gaps that form symmetrically around $\nu_{gc}$ depend on $p_{max}$.  To illustrate this, consider once again the integer states $\nu=m$.  These will be part of all sequences explored through steps of $1/(2p+1)$, with $p\le p_{max}$.  Since the minimum step is $\Delta \nu=1/(2p_{max}+1)$, it follows that for a sufficient $n_{max}$, the state at $\nu_{gc}$ will be at the center of a $2\Delta \nu$ gap, as found in the example reported in Fig.\ref{Figure_number_sequence}c (top).  As for the gap-state of Fig.\ref{Figure_number_sequence}b, so also the fermionic gap-state is accompanied by its associated gap-center state in the bosonic spectrum, as shown in Fig.\ref{Figure_number_sequence}c (bottom). 

\subsubsection*{Spectrum self-similarity}

Another fundamental property of the spectrum described by Eqs.(\ref{fermions}) and (\ref{bosons})  is self-similarity.  In fact, for the fermionic spectrum of Eq.(\ref{fermions}), for any $\nu$ in the spectrum there will always be a $\nu^*=\nu/(2l+1)$ for $l$ integer, as long as the condition $(2p^*+1)=(2p+1)(2l+1)$ can be satisfied.  The latter condition implies that self-similarity is ultimately limited by $p_{max}$.  An example of this is reported in Fig.\ref{Figure_number_sequence}d for $n_{max}=20$, $p_{max}=20$, and $l=1$, for the interval $0<\nu<1$.  Evidently, while the overall features follow a self-similar behavior, the details, such as the widths of the gaps and gap-center states, will not be exactly self-similar.  To oberve the full mathematical realization of the self-similarity would require a larger $p^*_{max}$  (i.e., $(2p^*_{max}+1)=(2p_{max}+1)(2l+1)$).  An analogous self-similarity occurs for the bosonic spectrum of Eq.(\ref{bosons}),  for $\nu^*=\nu/l$, with $l$ integer, as long as the condition $n^*q^*/p^*=nq/pl$ is satisfied.

Self-similarity  allows us to formulate  general statements on the expected values of $\nu_{g}$ and $\nu_{gc}$. For example,  $\nu_g=1/2$ implies the set of $\nu_g^*=(1/2)/(2l+1)$, this including 1/6, 1/10, and so on (the gaps at 1/6 and 1/10 are visible in Fig.\ref{Figure_number_sequence}d).  The $\nu_g=1/4$ implies $\nu_g^*=(1/4)/(2l+1)$, this including 1/12, 1/20, and so on. For the all-important gap-center states,  $\nu_{gc}=1$ leads to the self-similar gap-center states at 1/3, 1/5, and so on, $\nu_{gc}=2$ leads to 2/3, 2/5, and so on, $\nu_{gc}=3$ leads to 3/5, 3/7, and so on, $\nu_{gc}=4$ leads to 4/3, 4/5, 4/7, 4/9, and so on, $\nu_{gc}=5$ leads to 5/3, 5/7, 5/9, and so on, $\nu_{gc}=6$ leads to 6/5, 6/7, and so on, and $\nu_{gc}=7$ leads to 7/3, 7/5, and so on.

\subsubsection*{Fractional-integer spectrum correspondence}



Eq.(\ref{fermions}) for the fermionic spectrum naturally contains a correspondence between fractional states and those of the integer effect for an appropriately shifted magnetic field $H_0^*=H_0-H_0'$. Indeed,  since $G_{xz}=C/H_0$, with $C=\rho c$, $\rho$ being the sample 2D charge density \cite{Laughlin1998}, $\nu=(h/e^2)C/H_0=H_0(1)/H_0$ and $\nu^*=H_0(1)/(H_0-H_0')$, where $H_0(1)=(h/e^2)C$ is the magnetic field for which $\nu=1$. Imposing $\nu^*=n$ with $\nu=n(2q+1)/(2p+1)$ (Eq.(\ref{fermions})) requires that $H'_0/H_0(1)=(2p+1)/(n(2q+1))-1/n$, an equality that holds only if $q=0$ and $p=nr$, where $r$ is an integer.  Consequently, the correspondence holds for $H'_0/H_0(1)=2r$, $\nu=n/(2nr+1)$, and $\nu^*=n$. In particular, for $r=1$, we have the typically reported $H'_0/H_0(1)=2$, $\nu=n/(2n+1)$, and $\nu^*=n$ (see for example, Ref.\cite{Lin2014}), while for $r=2$ we have $H'_0/H_0(1)=4$, $\nu=n/(4n+1)$, and $\nu^*=n$. An analogous similarity occurs for the mirror reflected $\nu^*=-n$, in which case each state of  spectrum $-n$ is superimposed with the state $n/(2nr-1)$, a similarity typically reported for $r=1$ \cite{Lin2014}.  The similarity does not, in turn, hold for the bosonic spectrum described by Eq.(\ref{bosons}), since an equivalent superposition of the integer states with fractional ones would imply the condition $1/n=p/nq-H_0'/H_0(1)$, that holds for any $n$ only if $p=q$, i.e., for the integer states themselves.

\subsubsection*{Hall conductance}

\begin{figure}[t]
\centering
\includegraphics[width=1.0\columnwidth]{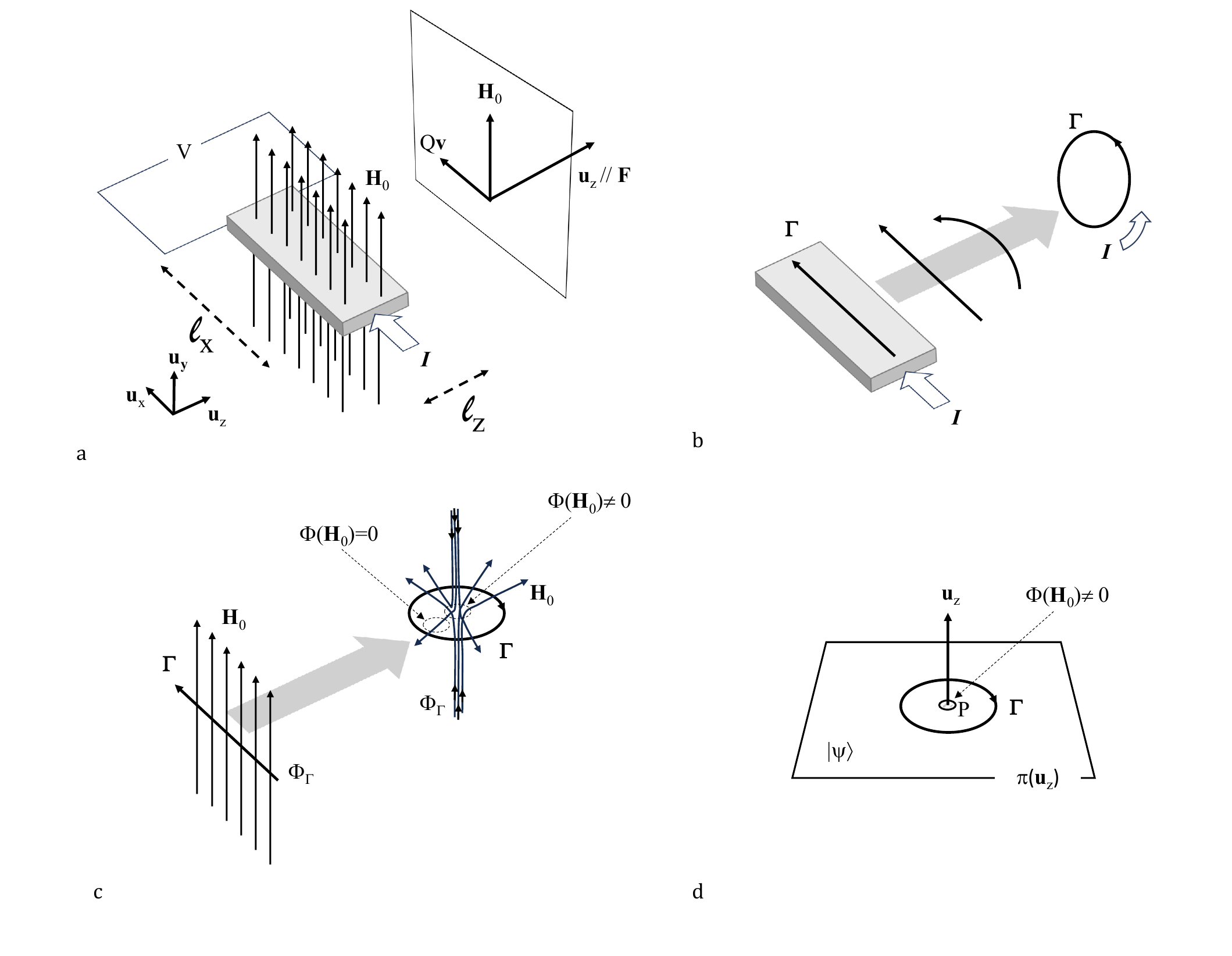}
\caption{\textbf{Laughlin loop.}  (a) Hall experiment geometry and the plane identified by the unit vector $\mathbf{u}_z$ parallel to the direction of the electromotive Hall force $\mathbf{F}=(Q/c)\mathbf{v}\wedge \mathbf{H}_0$, where $Q=-e$ is the electron charge and $\mathbf{v}$ its drift velocity. (b) Transformation of $\Gamma$ into a loop and (c) associated transformation of $\mathbf{H}_0$ and linked flux $\Phi_\Gamma$. (d) Recasting of the closed loop $\Gamma$ on the plane $\pi (\mathbf{u}_z)$ normal to $\mathbf{u}_z$. In terms of the Aharonov-Bohm effect, the forbidden point $P$ is  the  region of out-of-plane $\mathbf{H}_0$ with a finite  $\Phi \neq 0$, while all other regions, including $\Gamma$ itself, contribute with a zero linked field.}
\label{Figure_Laughlin_Loop}
\end{figure}

\begin{figure}[t]
\centering
\includegraphics[width=1.0\columnwidth]{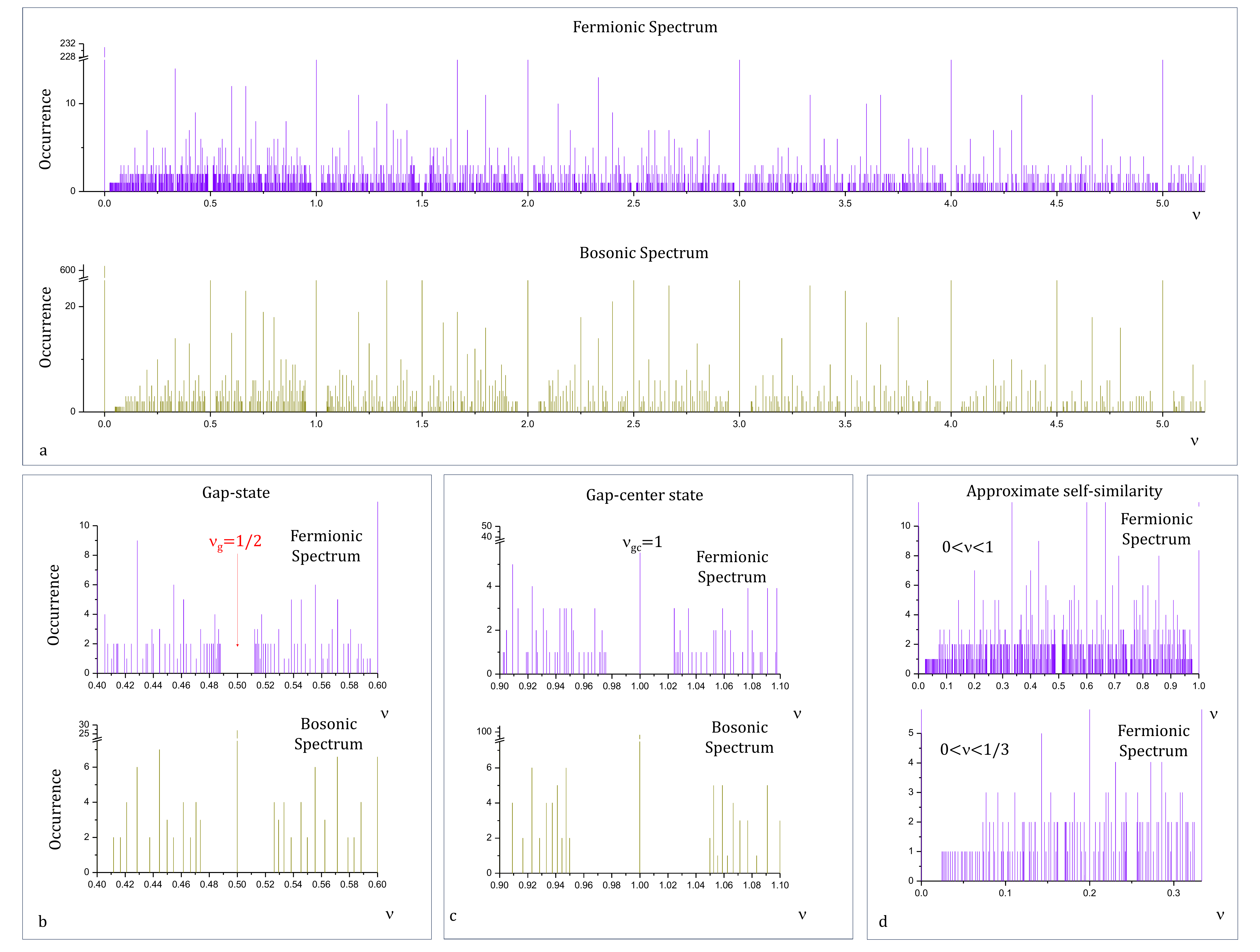}
\caption{\textbf{Spectrum sequence.} (a) Illustration of the number sequence of Eq.(\ref{fermions}) relative to fermionic part (top) and bosonic part (bottom) of the Hall spectrum for $n_{max}=20$ and $p_{max}=20$.  (b) Zoom into the gap in the fermionic spectrum at $\nu_g=1/2$ and (c) the gap-center state at $\nu_{gc}=1$.  Illustration (d) of the approximate self-similarity for the fermionic spectrum as expressed by $\nu'=\nu/(2l+1)$ for $l=1$ in the spectral region $0<\nu<1$ (top panel) and $0<\nu'<1/3$ (bottom panel).   Occurence is the number of different available combinations of $q$ and $p$ that lead to the same $\nu$.}
\label{Figure_number_sequence}
\end{figure}

\begin{figure}[t]
\centering
\includegraphics[width=1.0\columnwidth]{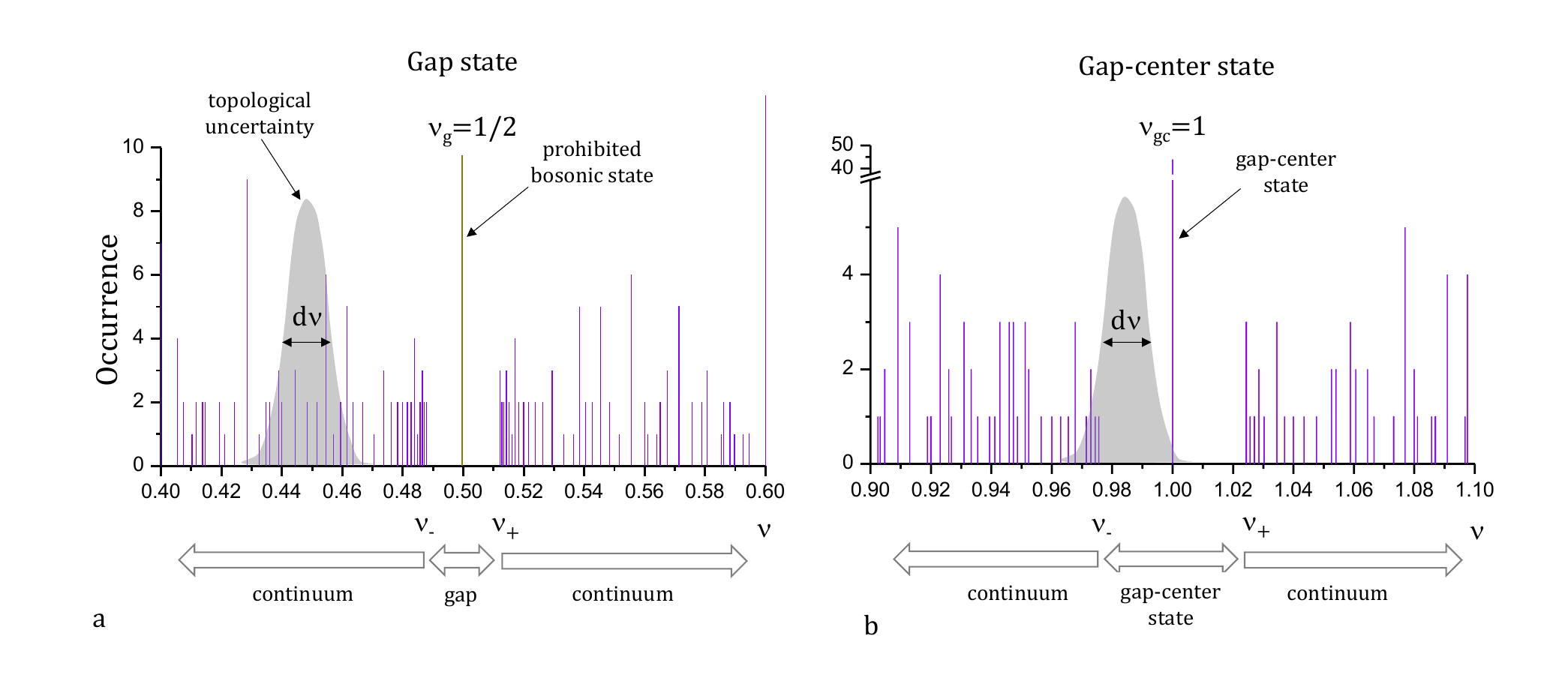}
\caption{\textbf{Origin of plateaus in proximity of gap-center states at $\nu_{gc}$.}  The intrinsic uncertainty in $\nu$ means that for any given value of applied $H_0$, the actual $\nu$ experienced by the electrons has a quantum spread $\delta \nu$ (shaded area).  (a) When the scan encounters a gap, the presence of a prohibited bosonic extended state causes transport to be classical.  (b) For the gap-center state, transport will be pinned to the isolated gap-center-state, fixing the value of  $G_{xz}=\nu_{gc}(e^2/h)$ for the finite set of values of $\nu_-<\nu<\nu_+$, depending on $\delta \nu$.}
\label{Figure_origin_of_plateaus}
\end{figure}

\noindent \textit{Role of $\Delta \nu$}. We next proceed to translate Eqs.(\ref{fermions}) and (\ref{bosons}) into predictions on measurements of $G_{xz}$ versus applied magnetic field $H_0$.  As illustrated in Fig.\ref{Figure_Laughlin_Loop}, changing $H_0$ changes the flux through the sample surface that, according to the Laughlin loop geometry and flux conservation, is the linked flux $\Phi_\Gamma$.  The magnetic field will lead to the $x$-delocalized Landau Levels (LLs) and edge modes on condition that wavefunction regularity is satisfied, i.e., if the resulting $\nu$  obeys Eq.(\ref{fermions}) or Eq.(\ref{bosons}).  When regularity cannot be ensured, the levels cannot form and the system will behave as a standard conductor in the presence of a magnetic field.   To determine the effect this mechanism has on electronic transport, we need to consider the implications of the underlying superposition of Eq.(\ref{superposition}).  What we have termed minimum step $\Delta \nu$ in the sequence of allowed values of $\nu$ in the fermionic spectrum is directly associated to the uncertainty in the topological angular momentum through the value of $j_z^{max}$,  $\Delta \nu=1/2j_z^{max}$.   Put differently, $\Delta \nu$ is not simply the minimum step in the sequence of allowed fermionic filling factors, but it is more precisely proportional to the minimum uncertainty in the value of $\nu$ associated to the quantum uncertainty in $J_z$, an uncertainty that is fixed by how strong the electrons are confined to the 2D metal system. Analogously, the minimum step in the corresponding bosonic part of the spectrum (i.e., from Eq.(\ref{bosons})) is $\Delta \nu'=1/p_{max}=1/(j_z^{max}-1/2) \simeq 2\Delta \nu$ (for $j_z^{max}>1/2$).  Hence, even for a fixed number of conduction electrons $N_e$ and applied field $H_0$, yet the value of $\nu$ has a finite minimum uncertainty $\delta \nu$ proportional to $\Delta \nu$ (to simplify the discussion, we can assume for now $\delta \nu=\Delta \nu$).   Consider the sequence reported in Fig.\ref{Figure_number_sequence}a.  The gap and gap-center structures recalled in Fig.\ref{Figure_number_sequence}b and Fig.\ref{Figure_number_sequence}c then become paradigmatic, as the gap has a width of $\Delta \nu$, $\Delta \nu$ that is also the distance of the gap-center state from the rest of the spectrum.  Assuming then that we scan experimentally growing values of $\nu$ for the spectrum of Fig.\ref{Figure_number_sequence}b, the states below the gap form a continuum, as their features are smaller than $\delta \nu$, meaning that the observed behavior is expected to follow the average classical behavior $G_{xz}=C/H_0$ (see Fig.\ref{Figure_origin_of_plateaus}a).  The same can be said of values of $\nu$ above the gap.  Values within the gap, in turn, correspond to a totally different situation. In the gap, a sole isolated bosonic state is present, the bosonic equivalent of the fermionic gap-center state of Fig.\ref{Figure_number_sequence}c.  So, while $\delta \nu$ can potentially cause electrons to span the gap, delocalized behavior is prohibited by the presence of the isolated boson state.  This will once again cause the system to manifest classical transport properties, that now exactly follows the classical behavior $G_{xz}=C/H_0$, not because $\delta \nu$ causes an average behavior on allowed delocalized states, but because delocalized behavior is absent.  Hence, in terms of $G_{xz}$, the quantum nature of the gap is not directly observable.  Of course, the gap can be observed analyzing $G_{xx}$, i.e., the longitudinal conductance.  Here, behavior associated to the continuum above and below the gap can give rise to fluctuations as the system continuously passes from electron transport associated to delocalized states with negligible resistance and classical states with finite resistance.  In turn, the values of $\nu$ in the gap will give rise to an actual classical behavior, i.e., that associated to a standard conductor manifesting magneto-resistance.  A very different scenario emerges for the gap-center state reported in Fig.\ref{Figure_number_sequence}c. Once again, values of $\nu$ above and below the structure form a continuum, as discussed above.  In turn, values within the gap-center structure will extend to the single isolated and degenerate gap-center state, so that there will be a finite portion of the spectrum where the system will manifest transport properties associated with a single isolated state (see Fig.\ref{Figure_origin_of_plateaus}b).  In terms of $G_{xz}$, this will lead to a plateau at the value $G_{xz}=\nu_{gc}(e^2/h)$, strongly deviating from the averaged $G_{xz}=C/H_0$. It goes without saying that an analogous anomaly will intervene in $G_{xx}$, where a negligible resistance will be observed in correspondence of the plateau, as electron behavior will be dominated by a single delocalized state.

\noindent \textit{Scanning versus magnetic field and carrier density}. In a large portion of experiments, the spectrum is achieved scanning $H_0$, not $\nu$. Starting from the average scaling $G_{xz}= C/H_0$ paired with Eq.(\ref{fundamental}),  the $\delta \nu$ uncertainty translates into an effective uncertainty in the magnetic field experienced by the electrons, with a spread $\delta H_0  \simeq (H_0^2/C) (e^2/h) \delta \nu$ that grows quadratically in $H_0$.  The implication is that, on analyzing experimental results, we should consider that the quantum uncertainty grows quadratically in magnetic field, so what we consider a continuum  or a gap-center state depends on the actual value of $H_0$. This said, for a given set of explored values  $H_0\le H_0^{max}$, the features of the spectrum are  fixed, through $\Delta \nu$,  by the finite value of $p_{max}$, which is fixed by the maximum available  $j^{max}_{z}=p_{max}+1/2$ for electrons (Eq.(\ref{fermions})), and $j^{max}_{z}=p_{max}$ for Cooper-pairs (Eq.(\ref{bosons})). For values of $H_0$ that correspond to the gaps at $\nu_{g}$ no extended states are present in the spectrum  so that  the material behaves like a standard metal, $R_{xz}$ increasing in proportion to $H_0$ and $R_{xx}$ being approximately constant, apart from the inevitable changes associated to standard magnetoresistance. In turn, since all values of $H_0$ that correspond to values of $\nu$ that  fall inside the gap-center spectrum have a fixed $R_{xz}=(\nu_{gc}e^2/h)^{-1}$, a plateau will form whose width $2\Delta H_0$ then provides an estimate of $p_{max}$ for the given sample. This is because, for $m$ integer, the plateau forms from $\nu_-=m-1/(2p_{max}+1)$ to $\nu_+=m+1/(2p_{max}+1)$, so that $2\Delta H_0/H_0(1)=(2/(2p_{max}+1))(m^2-1/(2p_{max}+1)^2)^{-1}$ ($p_{max}\ge 1$).  In turn, for these values of $H_0$, since the conducting electrons are pinned to the highly degenerate extended gap-center state, $R_{xx}=0$.

Measurements of $R_{xx}$ and $R_{xz}$ performed changing the charge density $\rho$ at a fixed $H_0$, in turn, amount to an actual scan in $\nu$.  This follows from $G_{xz}= C/H_0$, where $C=\rho c$, so that, from Eq.(\ref{fundamental}), $\rho$ and $\nu$ are, on average, proportional.  Plateau formation is then the result of topological uncertainty through the associated uncertainty $\delta \rho=(e^2/hc)H_0\delta \nu$,  the analysis proceeding in the same manner as for $H_0$ scans, the starting point being the evaluation of $p_{max}$.    

\noindent \textit{Spin-splitting: Altering $J_z$ uncertainty}. The key role played by $\Delta \nu$ in determining the width of the plateaus means that these can be affected by any process that alters the superposition of Eq.(\ref{superposition}). Consider then the intrinsic LL spin-splitting associated to the Zeeman effect \cite{Englert1982}. Assuming a Zeeman energy split $\Delta E=g^*\mu_B H_0$, where $g^*$ is the electron effective g-factor and  $\mu_B$ is the Bohr magneton, spin-splitting will be able to affect the indistinguishablity of the electron spin only if it is noticeably larger than the effective Zeeman uncertainty associated to $\delta H_0$, i.e.,  $g^*\mu_B\delta H_0$.  This means that the splitting will be important in the seemingly counterintuitive situation of large values of $\nu$.  For example, a reasonable condition $H_0>2\delta H_0$ leads to the approximate condition that $\nu \ge 2$ for a $\delta \nu =1$.  In these conditions,  splitting will destroy the uncertainty in the electron spin along $\mathbf{u}_y$,   the direction of the applied field $\mathbf{H}_0$ (see Fig.\ref{Figure_Laughlin_Loop}a). In the Laughlin loop, this component is mapped radially in the  $\pi (\mathbf{u}_z)$ plane, the original +1/2 component mapped into a radially -out- component, the -1/2 mapped into the radially -in- component.  In terms of the topological angular momentum $\mathbf{J}$, these two components correspond to the $S_z=\pm 1/2$ states (a rotation around $\mathbf{u}_z$ in the abstract Laughlin loop space of Fig.\ref{Figure_Laughlin_Loop}d corresponds in the actual space of Fig.\ref{Figure_Laughlin_Loop}a to a translation along $\mathbf{u}_x$, so it cannot alter the state of the spin along $\mathbf{u}_y$).  By removing the spin component uncertainty,  spin-splitting will affect the uncertainty in $J_z$ in proportion to $1/j_z^{max}=1/(p_{max}+1/2)$,  leading to observable effects in $\delta \nu$ only for small values of $p_{max}$ (and, as discussed above, large values of $\nu$).   

\begin{figure}[t]
\centering
\includegraphics[width=1.0\columnwidth]{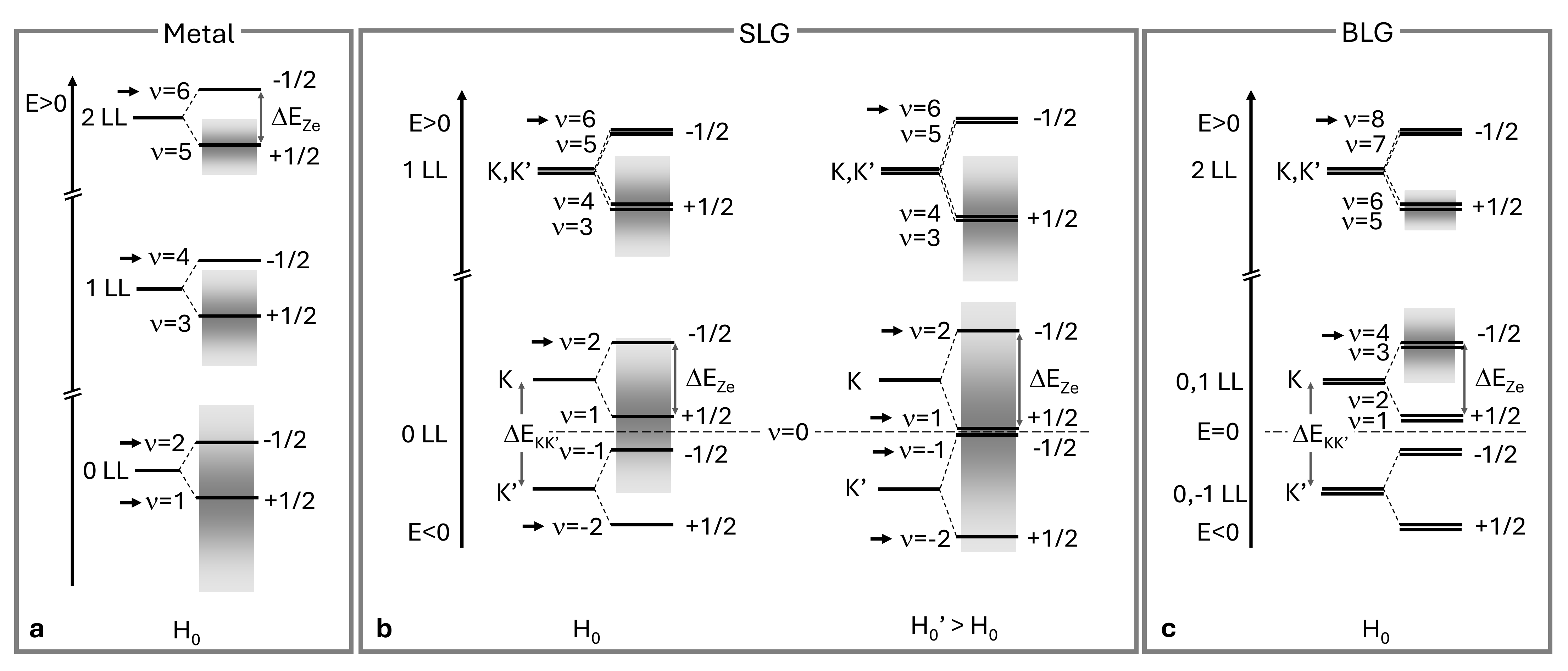}
\caption{\textbf{Zeeman splitting for $p_{max}=0$ and the role of LL spectrum.}  (a) For a standard metal system, topological uncertainty associated to $\Delta \nu$ (grey shaded bars) compared to Zeeman splitting ($\Delta E_{Ze}$) causes $\nu=1,2,4,6,...$ (but not $\nu=3,5,7,...$) to form unaltered extended plateaus (arrows).  (b) In single-layer graphene (SLG) the interplay also with valley-splitting ($\Delta E_{KK'}$) causes different plateau structures for different $H_0$.  Relatively low fields (left) will allow plateaus to form only at $\nu=\pm2,\pm6,\pm10,...$, while stronger fields (right) cause $\nu=\pm1,\pm2,\pm6,\pm10,...$ to form plateaus accompanied by a pseudo-plateau at $\nu=0$. (c) For bi-layer graphene (BLG) in conditions corresponding to SLG panel (b) left, $\nu=\pm4,\pm8,\pm12,...$ will form extended plateaus.}
\label{Figure_Zeeman_splitting_Graphene}
\end{figure}

\begin{figure}[t]
\centering
\includegraphics[width=1.0\columnwidth]{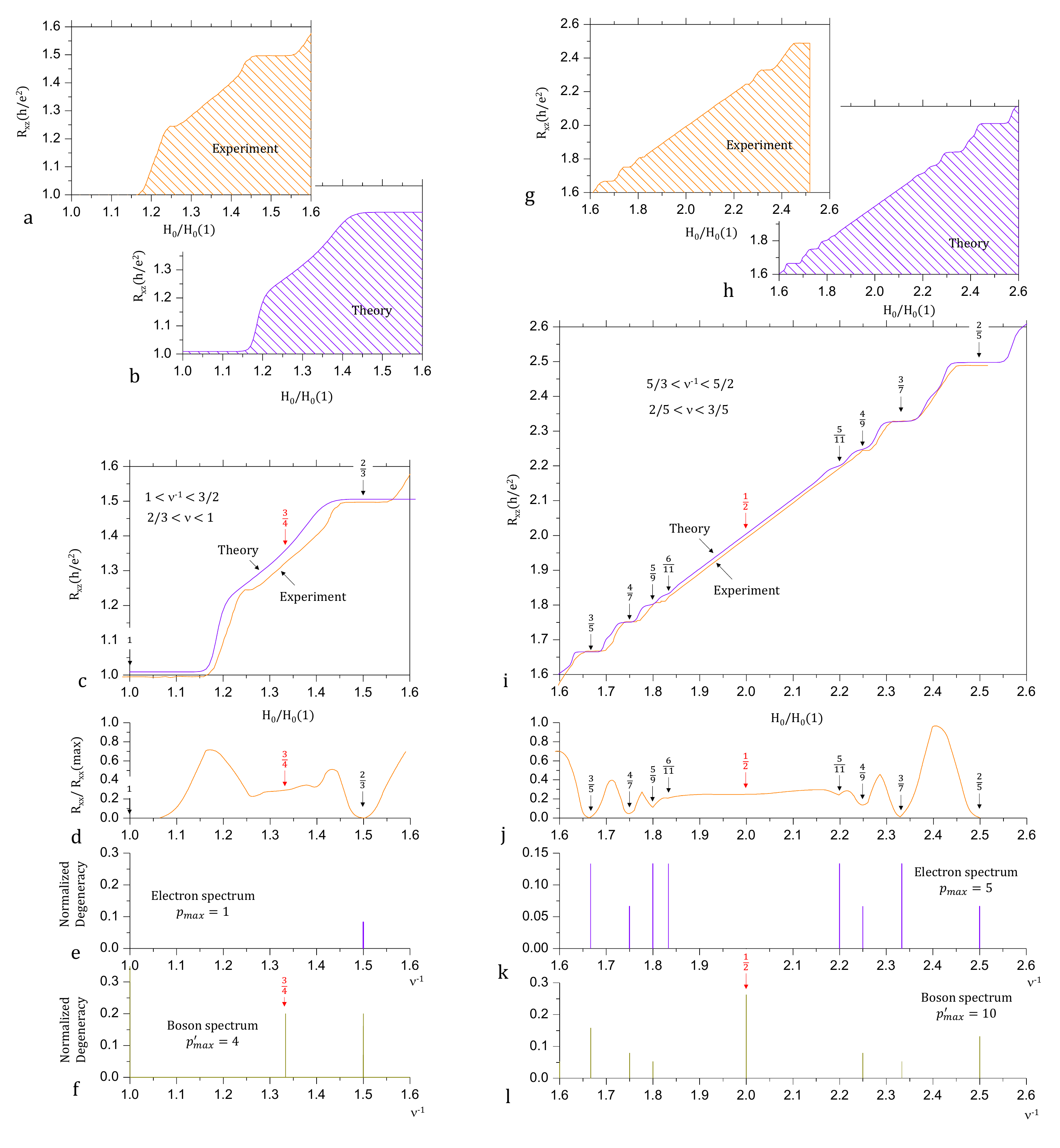}
\caption{\textbf{Further theory versus experiment in the FQHE ($2/3<\nu<1$).} (a) Hall resistance  measurements taken from Ref.\cite{Willett1987} (orange curve) for $2/3<\nu<1$ ($1<\nu^{-1}<3/2$) compared to (b) prediction based on Eqs.(\ref{fermions}) and (\ref{bosons}) and corresponding Eq.(\ref{Gxz_fermionic}) (violet curve), and (c) superimposed comparison.   (d) Observed normalized $R_{xx}$. (e) Gap-center states $\nu_{gc}$ and gaps $\nu_g$  are referenced back to the fermionic spectrum ($p_{max}=1$) and (f) to the corresponding topologically prohibited states of the bosonic spectrum ($p'_{max}=4$). (g)-(l) Analogous comparison for $2/5<\nu<3/5$ ($5/3<\nu^{-1}<5/2$). Note the change in $p_{max}$ for low (a)-(f) and high $H_0$ (g)-(l) scans (see text).}
\label{experiments_vs_theory_fractional_second}
\end{figure}

\noindent The spin-dominated $p_{max}=0$ case: The most important example of the effects of spin-splitting on plateaus is then the spin-dominated case of $p_{max}=0$ that  intervenes for sufficiently weak electron confinement to the sample plane.  For $p_{max}=0$,  $j=j_z^{max}=1/2$, and from Eq.(\ref{fermions}), $\nu=n$.  The implication is that not only are the allowed delocalized electron states always associated to an integer $\nu$, but they always coincide with exactly full LLs, a generally rare circumstance that is here the dominant feature of the spectrum.   More precisely,  $\Delta \nu$=1, and the spectrum necessarily only contains the integer values that belong to the subset of $\nu_{gc}$. There being no continuum of states and no corresponding meaningful $\nu_-$ and $\nu_+$,  electron transport is determined by the direct passage  from one extended state to the next, with plateaus that, in the absence of spin-splitting, would extend for each of the integer values from $\nu-1/2$ to the following  $\nu+1/2$, leading to a characteristic non-classical step-like behavior in $G_{xz}$ where no spectrum continuum or gaps intervene.  For $G_{xz}$ versus $H_0$, we expect these plateaus to have a width of $2\delta H_0/H_0(1)\simeq 2(H_0/H_0(1))^2$.  When spin-splitting intervenes for sufficiently high values of $\nu$, the picture is fundamentally altered.  To clarify this,  we begin by recalling that in the Laughlin loop scheme of Fig.\ref{Figure_Laughlin_Loop}, changes in the linked flux $\Phi_\Gamma$ compatible with regularity amount to gauge transformations for the electrons in the metal. Given the constraints associated to single-valuedness, this causes electron LLs to shift in the $\mathbf{u}_z$ direction, remapping all occupied states into themselves when a flux quantum $\delta \Phi$ is inserted through the loop \cite{Laughlin1981,Laughlin1999}.  For a finite sample of length $\ell_z$, this only formal gauge transformation leads to the observable transfer of $n$ electrons from one side of the sample to the other along $\mathbf{u}_z$, as per the discussion leading to Eq.(\ref{HallConductance}), where  the value of $n$ is the number of occupied LLs. It follows that for odd values of $\nu$, only one Zeeman-split LL will be fully occupied, meaning that even though Hall measurements are not a measurement of electron energy, yet the electrons themselves in the highest occupied level will not be in a superposition of spin.  In this case, the topological angular momentum $J_z$ will be fully determined and $\delta \nu$ will diverge.  The result is that, for these electrons, the odd value gap-center states will not give rise to extended plateaus, so that Hall response will effectively occur directly from one even number integer state to the other as the value of $\nu$ reaches the odd state. Evidently, the overall behavior of $G_{xz}$ will depend both on the electrons in this highest occupied LL and in all the other occupied levels, for which the spin remains undetermined, i.e., which continue to have a finite $\delta \nu$.  Since, however, for the highest level uncertainty diverges, it is this reduced odd values plateaus behavior that becomes dominant.   For even values of $\nu$, in turn, spin superposition is maintained, as Hall measurements do not distinghuish between the sign of the topological quantum momentum, nor do they actually measure electron energy.  The situation is illustrated in Fig.\ref{Figure_Zeeman_splitting_Graphene}a.

\noindent Different Landau Spectra and Graphene: Dominant spin-splitting for $p_{max}=0$ highlights the fact that, even though experiments and theory have a universal trait, embodied experimentally by the  fractional values of flux quantum and, in our theory, by the overarching role of topological uncertainty through $p_{max}$,  the details of the Landau spectrum, that do not share this universal nature, can cause specific plateaus to be diminished, and even to not form.  

To illustrate this point, consider the interesting case of single-layer graphene (SLG, see Fig.\ref{Figure_Zeeman_splitting_Graphene}b) \cite{Novoselov2005,Novoselov2006,Zhang2006,Jiang2007}.  Since the lowest zero-energy LL $\nu=0$ generally has a sample-dependent broken valley degeneracy, spin-splitting causes a quadruplet to form with two spin-split states originating in the positive energy electron-spectrum and two in the negative energy hole spectrum \cite{Zhang2006,Jiang2007}.  It follows that the response of electrons (holes) that occupy $\nu=+1$ ($\nu=-1$) are not solely affected by the Zeeman split $\nu=+2$ ($\nu=-2$) state, but also by the neighboring $\nu=-1$ ($\nu=+1$) as, for magnetic fields $H_0$ used in experiments, typical valley-splitting \cite{Wang2015,Freitag2016} is comparable to Zeeman splitting \cite{Zhang2006}.  So it is that while the $\nu=1$ plateau is observable in a standard metal-like spectrum because electrons will occupy indistinguishably the $\nu=1$ and $\nu=2$ (indistinguishability is expected to be reduced only for $\nu>2$, as discussed above and illustrated in Fig.\ref{Figure_Zeeman_splitting_Graphene}a), in SLG, the superposition also involves the $\nu=-1$ level (see Fig.\ref{Figure_Zeeman_splitting_Graphene}b left), reducing the uncertainty in $J_z$ and hence causing the $\nu=1$ plateau to be greatly curtailed and even to disappear.  A signature of this more elaborate superposition of states is then that, for stronger values of $H_0$,  $\delta H_0$ will eventually span the entire quadruplet $\nu=-2,-1,1,2$, once again restoring the $J_z$ uncertainty and allowing a $\nu=1$ ($\nu=-1$) plateau to form (see Fig.\ref{Figure_Zeeman_splitting_Graphene}b right).  In this case, in turn, topological uncertainty implies that response will be dominated by a quantum superposition of electrons and holes, a superposition that then amounts to a rigid $G_{xz}=0$, a feature that appears in this respect similar to a $\nu=0$ plateau.  In truth, this is not an actual delocalized $\nu=0$ state (there is no state at $\nu=0$) but, rather, the superposition of different $\nu \neq 0$ states, meaning that, in distinction to standard plateaus, the $\nu=0$ plateau will not have its associated zero-resistivity region. Higher LLs associated to values of $\nu \ge 3$ have a four-fold valley and spin degeneracy.  For these, the Zeeman effect splits the valley-degenerate states into valley-degenerate spin-polarized pairs.  It follows that the $\nu=3,4,5$ states have a reduced $J_z$ uncertainty and, consequently, are not expected to manifest a plateau.  The result is that, for typical values of $H_0$, a sequence of plateaus $\nu=\pm 2, \pm 6, \pm 10, ...$ is expected (Fig.\ref{Figure_Zeeman_splitting_Graphene}b left).  Different conditions, such as stronger magnetic fields, are instead expected to lead to added plateaus at $\nu = \pm 1$ along with the formation of a pseudo-plateau at $\nu =0$ (as for SLG Fig.\ref{Figure_Zeeman_splitting_Graphene}b right).

Following an analogous  line of reasoning, a slightly different situation is expected to occur in BLG (see Fig.\ref{Figure_Zeeman_splitting_Graphene}c) where there is a further degeneracy in the zeroth and first LL \cite{McCann2006,Li2018}.  Here, for the lowest positive energy level formed by the degenerate valley-split zeroth and first LL and corresponding  highest negative energy (hole) level, and for all the other valley-degenerate electron and hole LLs, Zeeman splitting is expected to form pairs of doubly-degenerate spin-polarized states.  In conditions of intermediate field $H_0$ (i.e., those that lead to Fig.\ref{Figure_Zeeman_splitting_Graphene}b left), the $\nu=1,2$ plateaus are prohibited as for Fig.\ref{Figure_Zeeman_splitting_Graphene}b left, while the $\nu=3$ plateau is prohibited by the fact that the uncertainty $\delta H_0$ is smaller that $\Delta E_{Ze}$, as for the $\nu=3$ state in SLG.  In turn, the $\nu=4$ state now has the full uncertainty in spin (and hence $J_z$), and will manifest its plateau, and so for higher values of $\nu$.  In other words,  plateaus are expected for $\nu= \pm 4, \pm 8, \pm 12, ....$ (see Fig.\ref{Figure_Zeeman_splitting_Graphene}c). Evidently, as for SLG, conditions can be found in which the picture resembles the BLG version of Fig.\ref{Figure_Zeeman_splitting_Graphene}b right, in which case intermediate plateaus can be expected and even the $\nu=0$ pseudo-plateau \cite{Zhao2010}.

\noindent \textit{Temperature}. In order for extended states to lead to observable effects, a necessary condition is that thermal fluctuations must be smaller than the LL energy spacing, i.e., that, for a given $T$, $H_0\gg k_BTMc/\hbar e$, where $M$ is the effective electron mass (corresponding to fields on the order of 1 T or larger for $T\sim 1$K).  From the average relationship between $G_{xz}$ and $\nu$, the same reasoning also indicates that there will be a thermally-induced uncertainty in $\delta_T (1/\nu)\sim 2\pi k_BTM/h^2n_e \simeq 0.25 \cdot 10^{-2}$ (for $T \simeq 1$K and density of electrons $n_e \simeq 5 \cdot 10^{11}$ cm$^{-2}$). Hence, each allowed value of $\nu$ in the spectrum of Eq.(\ref{fermions}) and Eq.(\ref{bosons}) will have a finite resolvable width $\delta_T \nu = \delta_T (1/\nu) \nu^2$. Analogously,  the effects of spin-splitting will only be observed on condition that $k_BT$ be also smaller than the Zeeman energy split $\Delta E_{Ze}$ (and $\Delta E_{KK'}$ for SLG). It goes without saying that  to observe directly the spectrum of Eq.(\ref{bosons}), the generally stronger constraint that $k_BT$ be smaller than the Cooper pair gap must be met, so that at least a part of the electrons in the system behave as bosons \cite{Abrikosov1988}.  

\noindent \textit{Universal $G_{xz}$ plots}. For values of $T$ and $H_0^{max}$ that allow the observation of the  $\nu=\nu_{gc}=1$ plateau (at $R_{xz}=(e^2/h)^{-1}$),  the value of $C=\rho c$, that depends on the density of charge carriers of the specific sample, can be experimentally determined from the value of magnetic field  $H_0(1)$ at which the $\nu_{gc}=1$ plateau begins to develop as $T$ is lowered.  For a given $p_{max}>0$ and a sufficiently low $T$, universal curves will emerge plotting $R_{xz}$ versus $h_0=H_0/H_0(1)$ ($\nu=1/h_0$).  As discussed above, universality is broken for $p_{max}=0$, where spin-splitting depends on if we are considering 2D metals in semiconductor systems, SLG, BLG, or multi-layer graphene, and, in SLG, on the specific sample-dependent value of the valley-degeneracy split $\Delta E_{KK'}$ (see above). Non-universal behavior also holds for SLG for a generic $p_{max}>0$ at $\nu=0$, since here both electrons and holes can intervene in transport (see above).

\noindent \textit{Predicting $G_{xz}$}. The expected value of $G_{xz}$, as $\nu$ is scanned either by changing $H_0$ or $\rho$, depends on the set of delocalized fermionic states $\{ \nu_i \}$ with the corresponding set of occurences $\{N(\nu_i)\}$  (from Eq.(\ref{fermions})), and on the set of delocalized bosonic states $\{ \nu_i'\}$  and its $\{N(\nu_i')\}$ (Eq.(\ref{bosons})), this for the given $p_{max}$ and $n_{max}$.  The corresponding bosonic states are taken from the spectrum with $p_{max}' \simeq 2p_{max}+1$, assuming that the bosons are the consequence of two electrons forming a Cooper pair, with equivalent $j_{max}^z \simeq 2j_{max}^z$. If transport is  associated to single electrons and if $\nu=\nu_i$, then $G_{xz}/(e^2/h)=\nu_i$.  Moreover, if $\nu=\tilde{\nu}_i$ coincides with a value of the subset $\{ \tilde{\nu}_i'\}$ of bosonic states $\{ \nu_i'\}$  that are topologically prohibited delocalized states for fermions, the electrons cannot have a delocalized nature and transport is described by conventional magnetoresistance.  On consequence of topological uncertainty, i.e., of the finite value of $\Delta \nu$ determined by $p_{max}$, $G_{xz}$ will be the result of a superposition of these effects.  Assuming that topological uncertainty translates into a gaussian lineshape centered at $\nu$ in the Hall spectrum (see illustration in Fig.\ref{Figure_origin_of_plateaus}) of width $\delta \nu$ proportional to $\Delta \nu$, the contribution of each delocalized fermionic state to $G_{xz}$ will be $\nu_i$ multiplied by the probability that the system occupies the specific state, i.e., will be proportional to $\nu_iN(\nu_i)\exp{(-(\nu-\nu_i)^2/(\delta \nu)^2)}$.  Analogously, the contribution of the prohibited states will be $\nu$ multiplied by the probability that the system occupies the prohibited state, i.e., will be proportional to $\nu N(\tilde{\nu}_i')\exp{(-(\nu-\tilde{\nu}_i')^2/(\delta \nu')^2)}$, where now $\delta \nu'$ is the width of the bosonic lineshape, once again proportional to $\Delta \nu$.  Hence, for a system with solely electron transport,

\begin{equation}
\textstyle \frac{G_{xz}(\nu)}{e^2/h} =\frac{\sum\limits_{\{ \nu_i \}}e^{-\frac{(\nu-\nu_i)^2}{\delta \nu^2}}N(\nu_i)\nu_i+\sum\limits_{\{ \tilde{\nu}_i' \}}e^{-\frac{(\nu-\tilde{\nu}_i')^2}{\delta \nu'^2}}N(\tilde{\nu}_i')\nu}{\sum\limits_{\{ \nu_i \}}e^{-\frac{(\nu-\nu_i)^2}{\delta \nu^2}}N(\nu_i)+\sum\limits_{\{ \tilde{\nu}_i' \}}e^{-\frac{(\nu-\tilde{\nu}_i')^2}{\delta \nu'^2}}N(\tilde{\nu}_i')}.
\label{Gxz_fermionic}
\end{equation}

\noindent In turn, if transport is associated to bosons, as would occur for Cooper pairs, 

\begin{equation}
\textstyle \frac{G_{xz}(\nu)}{e^2/h} =\frac{\sum\limits_{\{ \nu_i' \}}e^{-\frac{(\nu-\nu_i')^2}{\delta \nu'^2}}N(\nu_i')\nu_i'}{\sum\limits_{\{ \nu_i' \}}e^{-\frac{(\nu-\nu_i')^2}{\delta \nu'^2}}N(\nu_i')},
\label{Gxz_bosonic}
\end{equation}

\noindent there being, in this case, no topologically prohibited states as any member of the set $\{\nu_i\}$ can be part of $\{\nu_i'\}$ for a sufficiently large $p_{max}$. For a given value of $T$, defining $\eta=\rho_s/\rho$ the ratio of the density of condensed electrons, i.e., that follow the bosonic spectrum associated to Cooper pairs, relative to total electron density, the expected value of $G_{xz}$ will be a weighted average, with $(1-\eta)$ electrons contributing through Eq.(\ref{Gxz_fermionic}) and $\eta$ electrons, now in the form of condensed Cooper pairs, contributing through Eq.(\ref{Gxz_bosonic}).

In this manner, $G_{xz}$ depends on a set of experimentally determined parameters.  The principal is $p_{max}$, that fixes the fermionic and bosonic spectrum and introduces the effects of spin-splitting for $p_{max}=0$.  For example,  an experiment manifesting the so-called IQHE is described by a $p_{max}=0$ and a $\Delta \nu=1$, while a spectrum that has a plateau at a $\nu=3/7$ must necessarily have a $p_{max} \ge 3$ ($2p_{max}+1=7$) for electrons.  If no features of the spectrum can be assigned values with a denominator larger than 7, then $p_{max}=3$.  A second parameter is the value of $\delta \nu$ that, while proportional to $\Delta \nu$ and hence fixed by $p_{max}$, depends on the details of the lineshape.   A third parameter is $\delta \nu'$ associated to the bosonic spectrum.  From Eq.(\ref{fermions}) and Eq.(\ref{bosons}) $p_{max}$ fixes the minimum step in the fermionic spectrum to $\Delta \nu=1/(2p_{max}+1)$ and for the bosonic spectrum to $\Delta \nu'=1/p_{max} \simeq 2\Delta \nu$.  It follows that, assuming lineshape to be the same for the fermionic and bosonic spectrum, $\delta \nu' \simeq 2 \delta \nu$. A fourth less critical experimental parameter is the integer $n_{max}$, i.e.,  the maximum value of $\nu$ achievable, and is determined, in  any one system, by the maximum LL that can be occupied.   Experiments where $p_{max}=0$ have the further complexity of involving in a dominant manner spin-splitting, meaning that predictions will depend on the nature of the system, i.e., if it is a semiconductor 2D metal, SLG, BLG, or a specific multi-layer graphene structure.  In this case, for a semiconductor 2D metal, an extra experimental parameter is the value $\nu_{ss}$ ($\ge 2$) at which spin-splitting intervenes and the associated reduced topological uncertainty $\delta \nu_{ss} < \delta \nu$.    Finally, experiments at ever lower values of $T$ will lead to the formation of bosonic Cooper pairs.  This  can be experimentally ascertained when plateaus begin to form at the prohibited fermionic values $\nu_g$ and  introduces a final temperature-dependent experimentally determined parameter, the ratio $\eta$.


\subsection*{Supplementary Text}

\subsubsection*{QHE: A modelling conundrum}

The phenomenological complexity, whereby a complete and universally accepted understanding is still lacking,  is well portrayed in a series of recent review articles and books \cite{Jain2014,Hansson2017,Halperin2000,Feldman2021}.  For example, the two principal microscopic theories, i.e., the so-called composite-fermion and Haldane-Halperin theories, make substantially different predictions \cite{Jain2014} and fail to provide exact statements on real physical systems where the FQHE has been realized (see Section III in \cite{Hansson2017}).  As J.K. Jain puts it in the opening section of Ref.\cite{Halperin2000}         
entitled ``The mystery of the fractional quantum Hall effect'', the theoretical conundrum consists in the fact that ``... at certain special filling factors nature conspires to eliminate the astronomical degeneracy to yield unique, non-degenerate ground states, which are certain entangled linear superpositions of all of the basis functions... This raises many questions. What is the organizing principle? What is the mechanism of the FQHE? What makes certain filling factors special? What is unique about the ground states at these fractions? What are their wave functions, and what physics do they represent? What are their excitations? What role does the spin degree of freedom play? What is the quantitative theory? How do gaps depend on the filling factor? What are the neutral collective modes and their dispersions? Finally, what other surprising phenomena lurk around the corner?” The research status is expressed in the opening sentences of Ref.\cite{Feldman2021}: ``...experiments on these systems continue to produce surprises, and the field of quantum Hall effects remains a vital area of condensed matter research today.'' 

The situation for electron transport in graphene is emblematic.  Sufficiently strong magnetic fields cause Shubnikov-de Haas oscillations to give way to the anomalous IQHE in single-layer-graphene (SLG) and to the unconventional IQHE in bilayer-graphene (BLG) \cite{Novoselov2004,Novoselov2005,Zhang2005,Novoselov2006}.  This is in agreement with the application of the IQHE gauge invariance non-interacting particle theory to the specific SLG/BLG spin-valley state degeneracy, an agreement that then becomes a strong confirmation of 2D Dirac fermion behavior \cite{CastroNeto2009}. The picture begins to get blurred for stronger magnetic fields\cite{Zhang2006,Jiang2007,Jiang2007_1}. Here an IQHE emerges that is incompatible with the expected spin-valley degeneracy, suggesting that many-body correlation may be at play, an explanation that retraces the many-body theory invoked for the FQHE in 2D metals, but now here invoked for an altered IQHE.  This, alongside evidence of an altogether different IQHE, incompatible not solely with degeneracy, but with the very basic phenomenological traits of the QHE, at the charge-neutral Dirac point. And then, as for the 2D metal case in semiconductor heterostructures, more refined SLG experiments  at strong magnetic fields allowed the observation of the FQHE in graphene, suggesting, in analogy to the theories developed for 2D metals, a many-body correlated electron picture\cite{Du2009,Dean2011,Ghahari2011,Feldman2012,Amet2015,Diankov2016,Kim2019,Kaur2024,Lu2024}.  The result is then that, in distinction to 2D metals where particle correlation is invoked in the FQHE but not in the IQHE, for graphene, novel many-body effects are invoked for a part of the IQHE and for the FQHE, this aside from the unexplained anomaly at the Dirac point.   As for the 2D metal model of the QHE in semiconductor heterostructures, so also for graphene the formulation of a consistent picture encounters an ever growing set of puzzles that require further experimental and theoretical efforts \cite{Kaur2024}.

\subsubsection*{Fixing $n_{max}$} 

We note that we have also fixed $n_{max}=160$.  While this setting plays a secondary role in the interpretation of data as compared to $p_{max}$, the latter fixing the minimum step in the spectrum $\Delta \nu$, it too is determined by the experimental conditions and is empirically estimated.  Observed values of integer $\nu=m$ will satisfy $n/m=(2p+1)/(2q+1) \ge 1$, meaning that $m \le n_{max}$.  It follows that $n_{max}$ can be measured as being the maximum value of detected integer $\nu$.  In truth, as clear in the results for $R_{xz}$ reported in Fig.\ref{experiments_versus_theory_integer} (top panel), the allowed values of $\nu$ for decreasing values of $H_0$ correspond to an ever more densely packed set, for a fixed experimental resolution in $H_0$, so that a clear plateau can be seen up to $m\simeq 10$, while higher values of $m$ are not easily identified, meaning that a direct measurement of $n_{max}$ is generally unfeasible.  That $n_{max}$ depends on the specific experiment can be appreciated by considering that in order to observe the QHE, the energy gap in the LLs $\hbar \omega_C \gg k_BT$, where $\omega_C=eH_0/Mc$, so that $H_0 \gg (2\pi k_BTMc/he)$.  Allowed changes in the linked flux $\Phi_\Gamma$ lead to an observable transfer of $n$ electrons from one side of the sample to the other, with  $n$ the number of occupied LLs.   Since the degeneracy of each level is $r \simeq AH_0/\delta \Phi \simeq N_e/n$, where $A=\ell_x \cdot \ell_z$ is the area of the sample, it follows that $n \ll n_eh^2/(2\pi k_BTM)$, since the largest flux quantum $\delta \Phi(j,j_z)$ is for $j=j_z$, with $n_e=N_e/A$. For typical conditions, i.e., $n_e \sim 5 \cdot 10^{11}$cm$^{-2}$, $M \simeq 0.07 m_e$, where $m_e$ is the electron mass, and $T \sim 0.5$K, $n \ll 10^3$, making $n_{max}=160$ an acceptable choice.

\subsubsection*{Anomalous and Unconventional QHE}

For relatively low values of $H_0$, the observed $G_{xy}$, with characteristic plateaus at $\nu=\pm 2, \pm 6, \pm 10,...$ is compatible with the $p_{max}=0$ case discussed in connection with Fig.  \ref{Figure_Zeeman_splitting_Graphene}b left \cite{Novoselov2004,Zhang2005,Novoselov2005,Zhang2006,Jiang2007,Jiang2007_1,Du2009,CastroNeto2009}.
Accordingly, in the same samples, higher values of $H_0$ allow the emergence also of other interger plateaus ($\nu=\pm 1,\pm 3,...$), along with the observation of the $\nu=0$ pseudo-plateau, as expected from the discussions leading to Fig.  \ref{Figure_Zeeman_splitting_Graphene}b right \cite{Zhang2006,Jiang2007,Jiang2007_1,Du2009}. As regards to the unconventional QHE observed in BLG \cite{Novoselov2006},   our theory with $p_{max}=0$ in conditions discussed for Fig.  \ref{Figure_Zeeman_splitting_Graphene}c, i.e., for relatvely low values of $H_0$, predicts the emergence of the observed $\nu=\pm 4, \pm 8, ....$ plateaus.  As for the SLG case, this picture is further validated by the appearance of intermediate integer value plateaus for higher values of $H_0$, along the same lines discussed in Fig. \ref{Figure_Zeeman_splitting_Graphene}b right \cite{Zhao2010}.

\subsubsection*{Non-interacting particles and anyons}

The quantum Hall effect remains to this day the principal experimentally accessible example of a topological state of matter, an intriguing phenomenon that has been observed in many different charge-carrier systems, these including new materials, such as graphene  \cite{Zhang2005,Novoselov2006}.  It is also considered as the potential playground and test-bed for an emerging applicative scenario, that of topological quantum computing \cite{Jacak2003,Ma2019}.  Key to this breakthrough is the ability to access topological order associated to ground-state degeneracy that can manifest both Abelian and non-Abelian fractional transformation properties and associated particle anyon-like statistics \cite{Dutta2022}.  According to present held beliefs, non-Abelian anyons arise as quasi-particles of many-body systems, topological solitons of highly correlated electrons.   While Abelian properties are built into the integer and the original fractional QHE Laughlin states, non-Abelian properties are thought to be possible in specific even-fractional states, such as the observed filling factor 5/2 state \cite{Park2020,Pu2023}.  The non-Abelian nature of these states emerges as the states themselves appear analogous to specific and well-known many-body entangled wavefunctions associated to quantum phase transitions in highly correlated electron systems \cite{Zeng2022}.  It follows that the identification of fractional QHE states with their non-Abelian entangled state properties is based on the physical idea that the fractional statistics and fractional charge is not of the elementary electrons, but rather of the quasi-particles that emerge out of strong interaction.  The involvement of many-body long range correlations implies that the entire field of the QHE and its possible applications in topological quantum computing is intimately tied to multi-particle interaction, a physics that is still little understood.  More specifically, if the fractional QHE is a many-body effect, it means that our ability to achieve a scalable quantum computer itself hinges on developing devices based on states that themselves require not only a non-perturbative control of the system, but more importantly, a scalable ability to control the interaction of ever-more particles.  This possibly unsurmountable hurdle spurns the development of new ideas where non-Abelian anyon behavior can arise in non-interacting electron systems \cite{Lunic2020}.   Here we have discussed how experiments in the QHE imply quantization in angular topological space, whereby fractional charge along with Abelian and non-Abelian anyons are the result of non-interacting quantum superposition, not interaction.       In other words, our present theory, which is the only one consistent with all experimental facts and explains the integer and fractional QHE, opens a new avenue to anyon physics, according to which fractional charge arises from a fractional flux quantum that is the consequence of quantization in angular topological space of single non-interacting particles or Cooper pairs. As such, the Abelian and non-Abelian properties should be the result of the basic commutation rules of a generalized topological angular momentum, without the need for many-body physics and complexity-driven dynamics.



\begin{thebibliography}{1}
%
%
%
%
\bibitem{Aharonov1959} Y. Aharonov and D. Bohm, Significance of Electromagnetic Potentials in the Quantum Theory, Phys. Rev. 115, 485-491 (1959).



\bibitem{Dirac1931} P. A. M. Dirac, Quantised singularities in the electromagnetic field, Proc. R. Soc. A \textbf{133} 60-72 (1931).

\bibitem{Golik2024} B. Golik, D. Jukic, and H. Buljan, Theory of Classical Electrodynamics with Topologically Quantized Singularities as Electric Charges, Laser Photonics Rev. \textbf{19}, 2400217 (2025).

\bibitem{London1950} F. London, \textit{Superfluids}  (John Wiley and Sons, New York, 1950), Vol. I, p. 152.

\bibitem{Laughlin1981} R. B. Laughlin, Quantized Hall conductivity in two dimensions, Phys. Rev. B 23, 5632-5633 (1981).

\bibitem{Klitzing1980} K. v. Klitzing, G. Dorda, and M. Pepper, New Method for High-Accuracy Determination of the Fine-Structure Constant Based on Quantized Hall Resistance, Phys. Rev. Lett. \textbf{45}, 494-497 (1980).

\bibitem{Tsui1982} D. C. Tsui, H. L. Stormer, and A. C. Gossard, Two-Dimensional Magnetotransport in the Extreme Quantum Limit, Phys. Rev. Lett. 48, 1559-1562 (1982).

\bibitem{Klitzing1985}  K. v. Klitzing, The Quantized Hall Effect, in \textit{Nobel Lectures, Physics 1981-1990} T. Frängsmyr, G. Ekspong (eds.) (World Scientific, Singapore, 1993).

\bibitem{Prange1990}  R. E. Prange and S. M. Girvin (eds.), \textit{The Quantum Hall effect}, 2nd ed., (Springer-Verlag, New York, 1990). 

\bibitem{Chakraborty1995} T. Chakraborty and P. Pietiläinen, \textit{The Quantum Hall Effects: Integral and Fractional} (Springer-Verlag Berlin, Heidelberg, 1995).

\bibitem{DasSarma1997} S. Das Sarma and A. Pinczuk (eds.), \textit{Perspectives in Quantum Hall Effects}  (Wiley, New York, 1997).

\bibitem{Stormer1999} Horst L. Stormer, Nobel Lecture: The fractional quantum Hall effect, Rev. Mod. Phys. 71, 875-889 (1999).

\bibitem{Jacak2003} L. Jacak, P. Sitko, K. Wieczorek, and A. Wójs, \textit{Quantum Hall Systems: Braid groups, composite fermions, and fractional charge} (Oxford University Press, Oxford, 2003).

\bibitem{Du2009} X. Du, I. Skachko, F. Duerr, A. Luican, and E. Y. Andrei, Fractional quantum Hall effect and insulating phase of Dirac electrons in graphene, Nature 462, 192-195 (2009).

\bibitem{Ghahari2011} F. Ghahari, Y. Zhao, P. Cadden-Zimansky, K. Bolotin, and P. Kim, Measurement of the $\nu = 1/3$ Fractional Quantum Hall Energy Gap in Suspended Graphene, Phys. Rev. Lett. 106, 046801 (2011).

\bibitem{Dean2011} C. R. Dean, A. F. Young, P. Cadden-Zimansky, L. Wang, H. Ren, K. Watanabe, T. Taniguchi, P. Kim, J. Hone, and K. L. Shepard, Multicomponent fractional quantum Hall effect in graphene, Nat. Phys. 7, 693-696 (2011).

\bibitem{Feldman2012}  B. E. Feldman, B. Krauss, J. H. Smet, A. Yacoby, Unconventional Sequence of Fractional Quantum Hall States in Suspended Graphene, Science 337, 1196-1199 (2012).

\bibitem{Amet2015} F. Amet, A. J. Bestwick, J. R. Williams, L. Balicas, K. Watanabe, T. Taniguchi, and D. Goldhaber-Gordon, Composite fermions and broken symmetries in graphene, Nat. Commun. \textbf{6}, 5838 (2015). 

\bibitem{Willett2013} R. L. Willett, C. Nayak, K. Shtengel, L. N. Pfeiffer, and K. W. West, Magnetic-Field-Tuned Aharonov-Bohm Oscillations and Evidence for Non-Abelian Anyons at $\nu=5/2$, Phys. Rev. Lett. 111, 186401 (2013).

\bibitem{Diankov2016} G. Diankov, C.-T. Liang, F. Amet, P. Gallagher, M. Lee, A. J. Bestwick, K. Tharratt, W. Coniglio, J. Jaroszynski, K. Watanabe, T. Taniguchi, and D. Goldhaber-Gordon, Robust fractional quantum Hall effect in the N=2 Landau level in bilayer graphene, Nat. Commun. \textbf{7}, 13908 (2016). 

\bibitem{Dean2015} C. R. Dean, Even denominators in odd places, Nat. Phys. 11, 298-299 (2015).


\bibitem{Kim2019} Y. Kim, A. C. Balram, T. Taniguchi, K.i Watanabe, J. K. Jain, and J. H. Smet, Even denominator fractional quantum Hall states in higher Landau levels of graphene, Nat. Phys. 15, 154-158 (2019).

\bibitem{Lu2024} Z. Lu, T. Han, Y. Yao, A. P. Reddy, J. Yang, J. Seo, K. Watanabe, T. Taniguchi, L. Fu, and L. Ju, Fractional quantum anomalous Hall effect in multilayer graphene, Nature 626, 759-764 (2024).

\bibitem{Kaur2024} S. Kaur, T. Chanda, K.R. Amin, D. Sahani, K. Watanabe, T. Taniguchi, U. Ghorau, Y. Gefen, G.J. Sreejith, and A. Bid, Universality of quantum phase transitions in the integer and fractional quantum Hall regimes, Nat. Commun. \textbf{15}, 8535 (2024).



\bibitem{Laughlin1983} R. B. Laughlin, Quantized motion of three two-dimensional electrons in a strong magnetic field, Phys. Rev. B 27, 3383-3389 (1983).

\bibitem{Laughlin1983_1} R. B. Laughlin, Anomalous Quantum Hall Effect: An Incompressible Quantum Fluid with Fractionally Charged Excitation, Phys. Rev. Lett. \textbf{50}, 1395-1398 (1983).

\bibitem{Laughlin1998} R. B. Laughlin, Fractional Quantization, Rev. Mod. Phys. \textbf{71} 863-874 (1999).

\bibitem{Jain2007} J. K. Jain, Composite Fermions (Cambridge University Press, New York, 2007).

\bibitem{Feldman2021} D. E. Feldman and B. I. Halperin, Fractional charge and fractional statistics in the quantum Hall effects, Rep. Prog. Phys. \textbf{84}, 076501 (2021).

\bibitem{Halperin2020} B. I. Halperin, J. K. Jain, Eds., \textit{Fractional Quantum Hall Effects: New Developments}  (World Scientific, Singapore, 2020).



\bibitem{Jain1989} J. K. Jain, Composite-Fermion Approach for the Fractional Quantum Hall Effect, Phys. Rev. Lett. \textbf{63}, 199-202 (1989).

\bibitem{Jiang1989} H. W. Jiang, H. L. Stormer, D. C. Tsui, L. N. Pfeiffer, and K. W. West, Transport anomalies in the lowest Landau level of two-dimensional electrons at half-filling, Phys. Rev. B. \textbf{40}, 12013-12016 (1989).

\bibitem{Kang1993} W. Kang, H. L. Stormer, L. N. Pfeiffer, K. W. Baldwin, and K. W. West, How real are composite fermions?, Phys. Rev. Lett. 71, 3850-3853 (1993).

\bibitem{Pan1999} W. Pan, J.-S. Xia, V. Shvarts, D. E. Adams, H. L. Stormer, D. C. Tsui, L. N. Pfeiffer, K. W. Baldwin, and K. W. West, Exact Quantization of the Even Denominator Fractional Quantum Hall State at $\nu =5/2$ Landau Level Filling Factor, Phys. Rev. Lett. \textbf{83}, 3530-3533 (1999).

\bibitem{Pan2002} W. Pan, H. L. Stormer, D. C. Tsui, L. N. Pfeiffer, K. W. Baldwin, and K. W. West, Transition from an Electron Solid to the Sequence of Fractional Quantum Hall States at Very Low Landau Level Filling Factor, Phys. Rev. Lett. \textbf{88}, 176802 (2002).

\bibitem{Martin2004} J. Martin, S. Ilani, B. Verdene, J. Smet, V. Umansky, D. Mahalu, D. Schuh, G. Abstreiter, and A. Yacoby, Localization of Fractionally Charged Quasi-Particles, Science \textbf{305}, 980-983 (2004).

\bibitem{Lin2014} X. Lin, R. Du, and X. Xie, Recent experimental progress of fractional quantum Hall effect: 5/2 filling state and graphene, National Science Review \textbf{1}, 564-579 (2014).


\bibitem{Willett1987} R. Willett, J. P. Eisenstein, H. L. Störmer, D. C. Tsui, A. C. Gossard, and J. H. English, Observation of an even-denominator quantum number in the fractional quantum Hall effect, Phys. Rev. Lett. \textbf{59}, 1776-1779 (1987). See also H. L. Stormer, Physica B \textbf{177} 401-408 (1992). 


\bibitem{Jain2014} J. K. Jain, A note contrasting two microscopic theories of the fractional quantum Hall effect, Indian J. Phys. 88, 915-929 (2014). 


\bibitem{Hansson2017} T. H. Hansson, M. Hermanns, S. H. Simon, and S. F. Viefers, Quantum Hall physics: Hierarchies and conformal field theory techniques, Rev. Mod. Phys. 89, 025005 (2017).

\bibitem{Halperin2000} B. I. Halperin and J. K. Jain (Eds.), Fractional Quantum Hall Effects: New Developments  (World Scientific, Singapore, 2020).  

\bibitem{Novoselov2004} K. S. Novoselov, A. K. Geim, S. V. Morozov, D. Jiang, Y. Zhang, S. V. Dubonos, I. V. Grigorieva, A. A. Firsov, Electric Field Effect in Atomically Thin Carbon Films, Science 306, 666-669 (2004).

\bibitem{Novoselov2005}  K. S. Novoselov, A. K. Geim, S. V. Morozov, D. Jiang, M. I. Katsnelson, I. V. Grigorieva, S. V. Dubonos, and A. A. Firsov, Two-dimensional gas of massless Dirac fermions in graphene, Nature 438, 197-200 (2005).

\bibitem{Zhang2005} Y. Zhang, Y.-W. Tan, H. L. Stormer, and P. Kim, Experimental observation of the quantum Hall effect and Berry's phase in graphene, Nature 438, 201-204 (2005).

\bibitem{Novoselov2006} K. S. Novoselov, E. McCann, S. V. Morozov, V. I. Fal’ko, M. I. Katsnelson, U. Zeitler, D. Jiang, F. Schedin, and A. K. Geim, Unconventional quantum Hall effect and Berry’s phase of $2\pi$ in bilayer graphene, Nat. Phys. 2, 177-180 (2006).


\bibitem{CastroNeto2009} A. H. Castro Neto, F. Guinea, N. M. R. Peres, K. S. Novoselov, and A. K. Geim, The electronic properties of graphene, Rev. Mod. Phys. 81, 109 (2009).


\bibitem{Zhang2006} Y. Zhang, Z. Jiang, J.P. Small, M.S. Purewal, Y.-W. Tan, M. Fazlollahi, J.D. Chudow, J. A. Jaszczak, H.L. Stormer, and P. Kim, Landau-Level Splitting in Graphene in High Magnetic Fields, Phys. Rev. Lett.  96, 136806 (2006).

\bibitem{Jiang2007} Z. Jiang, Y. Zhang, H. L. Stormer, and P. Kim, Quantum Hall States near the Charge-Neutral Dirac Point in Graphene, Phys. Rev. Lett. 99, 106802 (2007). 

\bibitem{Jiang2007_1} Z. Jianga, Y. Zhang, Y.-W. Tana, H.L. Stormer, and P. Kim, Quantum Hall effect in graphene, Sol. State. Comm. 143, 14-19 (2007).















\bibitem{Hasan2010} Z. Hasan, and C. L. Kane, Colloquium: topological insulators, Rev. Mod. Phys. 82, 3045-3067 (2010).

\bibitem{Qi2011} X.-L. Qi, and S.-C. Zhang, Topological insulators and superconductors, Rev. Mod. Phys. 83, 1057-1110 (2011).

\bibitem{Khanikaev2017} A. B. Khanikaev and G. Shvets, Two-dimensional topological photonics, Nat. Photon. 11, 763-773 (2017). 

\bibitem{Freedman2003} M. H. Freedman, A. Kitaev,  M. J. Larsen, and Z. H. Wang, Topological quantum computation, Bull. Amer. Math. Soc. 40, 31 (2003).

\bibitem{Bartolomei2020} H. Bartolomei, M. Kumar, R. Bisognin, A. Marguerite, J.-M. Berroir, E. Bocquillon, B. Plaçais, A. Cavanna, Q. Dong, U. Gennser, Y. Jin, and G. Fève, Fractional statistics in anyon collisions, Science 368, 173-177 (2020).













\bibitem{Park2020} J. Park, C. Spanslatt, Y. Gefen, and A. D. Mirlin, Noise on the non-Abelian $\nu =5/2$ Fractional Quantum Hall Edge, Phys. Rev. Lett. 125, 157702 (2020).


\bibitem{Pu2023} S. Pu, A. C. Balram, M. Fremling, A. Gromov, and Z. Papic, Signatures of Supersymmetry in the $\nu =5/2$ Fractional Quantum Hall Effect, Phys. Rev. Lett. 130, 176501 (2023).



\bibitem{Nayak2008} C. Nayak, S. H. Simon, A. Stern, M. Freedman, and S. D. Sarma, Non-Abelian anyons and topological quantum computation, Rev. Mod. Phys. 80, 1083 (2008).

\bibitem{Dutta2022} B. Dutta, W. Yang, R. Melcer, H. K. Kundu, M. Heiblum, V. Umansky,
Y. Oreg, A. Stern, and D. Mross, Distinguishing between non-abelian topological orders in a quantum Hall system, Science 375, 193-197 (2022).





\bibitem{Merzbacker1962} E. Merzbacher, Single Valuedness of Wave Functions, Am. J. of Phys. \textbf{30}, 237 (1962).

\bibitem{Hall2015} B. Hall, \textit{Lie Groups, Lie Algebras, and Representations}, 2nd Ed. (Springer, 2015) - page 101.

\bibitem{Cohen-Tannoudji1977} see, for example, C. Cohen-Tannoudji, B. Diu, and F. Laloë, \textit{Quantum Mechanics} (Wiley, New York, 1977) - page 287.




\bibitem{Englert1982} Th. Englert, D. C. Tsui, A. C. Gossard, and Ch. Uihlein, g-Factor Enhancement in the 2D Electron Gas in GaAs/AlGaAs Heterojunctions, Surf. Sci. \textbf{113}, 295-300 (1982).

\bibitem{Laughlin1999} R. B. Laughlin, Nobel Lecture: Fractional Quantization, Rev. Mod. Phys. \textbf{71}, 863 (1999).

\bibitem{Wang2015} W.-X. Wang, L.-J. Yin, J.-B. Qiao, T. Cai, S.-Y. Li, R.-F. Dou, J.-C. Nie, X. Wu, and L. He, Atomic resolution imaging of the two-component Dirac-Landau levels in a gapped graphene monolayer, Phys. Rev. B 92, 165420 (2015).

\bibitem{Freitag2016}   N. M. Freitag, L. A. Chizhova, P. Nemes-Incze, C. R. Woods, R. V. Gorbachev, Y. Cao, A. K. Geim, K. S. Novoselov, J. Burgdorfer, F. Libisch, and M. Morgenstern, Electrostatically Confined Monolayer Graphene Quantum Dots with Orbital and Valley Splittings, Nano Lett. 16, 5798-5805 (2016).

\bibitem{McCann2006} E. McCann, and V. Fal’ko, Landau Level Degeneracy and Quantum Hall Effect in a Graphite Bilayer, Pys. Rev. Lett. 96, 086805 (2006).

\bibitem{Li2018} J. Li, Y. Tupikov, K. Watanabe, T. Taniguchi, and J. Zhu, Effective Landau Level Diagram of Bilayer Graphene, Phys. Rev. Lett. 120, 047701 (2018).

\bibitem{Zhao2010} Y. Zhao, P. Cadden-Zimansky, Z. Jiang, and P. Kim, Symmetry Breaking in the Zero-Energy Landau Level in Bilayer Graphene, Phys. Rev. Lett. 104, 066801 (2010).



\bibitem{Abrikosov1988} A. A. Abrikosov, \textit{Fundamentals of the theory of metals} (Elsevier, Amsterdam,1988).


\bibitem{Paalanen1982} M. A. Paalanen, D. C. Tsui, and A. C. Gossard, Quantized Hall effect at low temperatures, Phys. Rev. B \textbf{25}, 5566-5569 (1982). 


\bibitem{Schmitz2020} M. Schmitz, T. Ouaj, Z. Winter, K. Rubi, K. Watanabe, T. Taniguchi, U. Zeitler, B. Beschoten, and C. Stampfer, Fractional quantum Hall effect in CVD-grown graphene, 2D Mater. 7, 041007 (2020).

\bibitem{Zibrov2018} A. A. Zibrov, E. M. Spanton, H. Zhou, C. Kometter, T. Taniguchi, K. Watanabe, and A. F. Young, Even-denominator fractional quantum Hall states at an isospin transition in monolayer graphene, Nat. Phys. \textbf{18}, 930-935 (2018).


\bibitem{Mani1996} R.G. Mani and K. von Klitzing, Fractional quantum Hall effects as an example of fractal geometry in nature, Z. Phys. B \textbf{100}, 635-642 (1996).







\bibitem{Ma2019} K. K. W. Ma and D. E. Feldman, The sixteenfold way and the quantum Hall effect at half-integer filling factors, Phys. Rev. B \textbf{100}, 035302 (2019).

\bibitem{Zeng2022} T.-S. Zeng and W. Zhu, Chern-number matrix of the non-Abelian spin-singlet fractional quantum Hall effect, Phys. Rev. B \textbf{105}, 125128 (2022).

\bibitem{Lunic2020} F. Lunic, M. Todoric, B. Klajn, T. Dubcek, D. Jukic, and H. Buljan,  Exact solutions of a model for synthetic anyons in a noninteracting system, Phys. Rev. B \textbf{101}, 115139 (2020).

\bibitem{methods} See Materials and Methods.





\end{thebibliography}
\end{document}